\documentclass{optica-article}
\journal{opticajournal}

\articletype{Research Article}

\usepackage{subfiles}
\usepackage[section]{placeins}
\usepackage{lineno}
\usepackage[version=4]{mhchem}
\usepackage{siunitx}
\mathchardef\mhyphen="2D
\definecolor{oceanboatblue}{rgb}{0.0, 0.47, 0.75}

\begin{document}
	\title{Entanglement signatures for quantum synchronization with single-ion phonon laser}
	
	\author{SI-WEN HE,\authormark{1,2} ZHI JIAO DENG,\authormark{1,2,3,*} YI XIE,\authormark{1,2} YAN-YI WANG,\authormark{1,2,4} and PING-XING CHEN\authormark{1,2}}
	
	\address{\authormark{1}Institute for Quantum Science and Technology, College of Sciences, National University of Defense Technology, Changsha, Hunan, 410073, China\\
		\authormark{2}Hunan Key Laboratory of Mechanism and Technology of Quantum lnformation, Changsha, Hunan, 410073, China\\
		\authormark{3}Key Laboratory of Low Dimensional Quantum Structures and Quantum Control of Ministry of Education, Hunan Normal University, Changsha, Hunan, 410081, China\\
		\authormark{4}College of Science, Guangxi University of Science and Technology, Liuzhou, Guangxi, 545006, China\\}
	
	\email{\authormark{*}dengzhijiao926@hotmail.com} 
	

	\begin{abstract*} 	
		The entanglement properties of quantum synchronization, based on a single-ion phonon laser subjected to an external drive, have been studied. It is found that the maximum value of steady-state entanglement between the ion's internal and external states occurs near the noiseless boundary from synchronization to unsynchronization, accompanied by noticeable oscillatory behaviors during the corresponding time evolution of entanglement. In addition, the later time dynamics of entanglement also indicates the occurrence of frequency entrainment, as evidenced by the strong consistency between the bending of the observed frequency and the emergence of Liouvillian exceptional points (LEPs) in the first two eigenvalues of the Liouvillian eigenspectrum. Moreover, the emergence of LEPs, which is intimately associated with frequency entrainment, should be widely observed in quantum synchronization and can be explored in LEPs-based applications.
	\end{abstract*}
	\section{Introduction}
	\label{sec:intro}
	Synchronization refers to the rhythmic adjustment of self-sustaining systems to weak disturbances and is prevalent in nature, encompassing the fields of physics, biology, chemistry, and engineering \cite{nonlinear}. While most research has traditionally focused on synchronization in classical systems, recent years have seen exploration of synchronization in the quantum regime with various quantum systems, including spins \cite{qs20180,qs2018,qs2020spin1,qs2023} or nonlinear oscillators \cite{qs2013prl,qs2014prl,qs2014tongue,qs2017pra,qs2018yasuo,qs2021coherence,qs2021jiadai} that rely on atomic ensembles \cite{qs2014atom,qs2016pra}, trapped ions \cite{qs2013prl,qs2015hush,qs2023} or optomechanical systems \cite{qs2015wigner,qs2016talita,qs2017gl}.
	
	In a trapped-ion system \cite{trappedion}, the generation of a self-sustaining oscillation of the ion's vibration, namely the single-ion phonon laser \cite{pl2009,pl2013} is typically based on the van der Pol (vdP) oscillator \cite{nonlinear}. It has been applied in injection locking \cite{pl2010} and the detection of weak forces \cite{pl2021}. To extend it to the quantum domain, incoherent sideband heating and cooling of the ion's vibrational motion can be employed \cite{qs2013prl}. Studies on quantum synchronization involve two scenarios: forced synchronization \cite{forced} of a single oscillator subjected to an external drive \cite{qs2013prl,qs2014prl,qs2017pra,qs2018yasuo,qs2021coherence,qs2021jiadai} and mutual synchronization between coupled oscillators \cite{qs2014tongue,qs2015huxinxi}. The synchronization of the quantum version is influenced by both quantum and thermal noise. Various approaches have been proposed to mitigate the impact of noise, such as the use of a squeezed drive \cite{qs2018yasuo,qs2021coherence,qs2021jiadai} and the implementation of measurement feedback control \cite{qs2021feedback}. Despite numerous theoretical investigations, to the best of our knowledge, no related experimental research has been reported yet, primarily due to the requirement of second-order sideband cooling to enable nonlinear dissipation \cite{qs2013prl}. 
	
	An alternative model for implementing a single-ion phonon laser in the quantum regime involves coherently driving the ion's vibrational motion with a blue sideband laser while simultaneously damping it with a cooling laser \cite{atomlaser,ionlaser,qs2015hush}. Hush et al. conducted a theoretical study on the synchronization of two such phonon lasers, revealing a strong correlation between the internal spin and external vibrational degrees of freedom within each ion \cite{qs2015hush}. They suggested that measuring the spin of individual ions could serve as an indirect method of confirming synchronization \cite{qs2015hush}. Quite similar to the phonon laser configuration mentioned above, the experimental realization of quantum synchronization with an external drive has been demonstrated in a trapped-ion system. In this experiment setup, two different ion species were used to manipulate the collective vibration, one for cooling and the other for heating to avoid cross talk \cite{pl2023}. On the other hand, quantum entanglement, which is a unique quantum property without a classical counterpart, has been studied in the context of quantum synchronization. The entanglement between two mutually synchronized systems has been investigated, aiming at revealing a relationship between entanglement and synchronization in coupled spins \cite{qs2018} or oscillators \cite{qs2014tongue}. Actually, there also exists entanglement within the above-mentioned experimentally feasible model, due to the fact that exciting the ion with a blue sideband laser is similar to the two-mode squeezing interaction used to generate entanglement \cite{gl2014}. Therefore, an intriguing question arises: Can the entanglement between the internal and external degrees of freedom indicate synchronization of a single-ion phonon laser to an external drive?
	
	In this paper, we study the entanglement properties of an experimentally feasible single-ion phonon laser with an external drive, with a specific focus on its relationship with quantum synchronization. First, the phase diagram of synchronization under the mean-field approximation for the noiseless case is presented. In this diagram, an explicit boundary separating synchronization from unsynchronization can be obtained. Based on this parameter phase diagram, it is found that the steady-state entanglement drops to zero in the deep synchronization region, whereas its maximum value appears near the noiseless boundary, with the corresponding time evolution of entanglement also exhibiting significant oscillatory behavior. After undergoing a transient process, the dynamics of entanglement is primarily determined by the first two eigenvalues of the Liouvillian eigenspectrum. In particular, the occurrence of Liouvillian exceptional points (LEPs) \cite{lep2019two,lep2019,lep2020pra,lep2020nori} in the first two eigenvalues is bound to happen and strongly linked to the frequency entrainment in quantum synchronization. This result sheds light on the characteristic changes in the Liouvillian eigenspectrum that we believe can be observed in all models of forced synchronization. The rest of this paper is organized as follows: In Sec. \ref{sec:model}, the system model and the parameter phase diagram of synchronization without noise are presented. In Sec. \ref{sec:ent}, the distribution of steady-state entanglement and its comparison with the quantum synchronization measure are studied. The dynamical evolution of entanglement, as well as the properties of the Liouvillian eigenspectrum, are analyzed in Sec. \ref{sec:dynamics}. The final part is left for the summary.
	
	\section{Model}
	\label{sec:model}
	In the laboratory frame, the total Hamiltonian $\hat{H}={{\hat{H}}_{0}}+{{\hat{H}}_{\operatorname{int}}}+{{\hat{H}}_{\text{ext}}}$ of the single-ion system is composed of three parts: the free Hamiltonian, the ion-laser interactions, and the external drive, which reads, respectively, ($\hbar =1$) \cite{tuidao,qs2015hush,exp1998}
	\begin{equation}
		{{{\hat{H}}}_{0}}=\frac{1}{2}{{\omega }_{0}}{{{\hat{\sigma }}}_{z}}+\nu {{{\hat{a}}}^{\dagger }}\hat{a},
	\end{equation}
	\begin{equation}
		{{{\hat{H}}}_{\operatorname{int}}}=\Omega{{{\hat{\sigma }}}_{+}}\left\{{{e}^{-i\left[ {{\omega }_{L}}t-\eta \left( {{{\hat{a}}}^{\dagger }}+\hat{a} \right) \right]}}+{{e}^{i\left[ {{\omega }_{L}}t-\eta \left( {{{\hat{a}}}^{\dagger }}+\hat{a} \right) \right]}} \right\} +h.c., 
	\end{equation}
	\begin{equation}
		{{{\hat{H}}}_{\text{ext}}}=iF\left( {{{\hat{a}}}^{\dagger }}{{e}^{-i\omega_{d} t}}-\hat{a}{{e}^{i\omega_{d} t}} \right), 
	\end{equation}
	where $\hat{a}$ $({{\hat{a}}^{\dagger}})$ is the annihilation (creation) operator for the ion's external vibration with vibrational frequency $\nu$. The ion's internal states involve the ground state $\left| g \right\rangle $ and excited state $\left| e \right\rangle $ with frequency difference ${{\omega}_{0}}$, and are described by Pauli matrix ${{\hat{\sigma }}}_{z}=\left| e \right\rangle \left\langle  e \right|-\left| g \right\rangle \left\langle  g \right|$ and ladder operators ${{\hat{\sigma }}}_{+}={{\hat{\sigma }}}_{-}^{\dagger}=\left| e \right\rangle \left\langle  g \right|$. The two internal states can be coupled with the external state by a laser with Rabi frequency $\Omega$, and the laser frequency is tuned to satisfy the first blue sideband resonance ${{\omega}_{L}}={{\omega}_{0}}+\nu$, for heating the ion's vibrational motion. Besides, $\eta$ is the Lamb-Dicke parameter and we assume that our system is in the Lamb-Dicke regime, defined by the condition $\eta^2(2 n+1) \ll 1$ and $n$ is vibrational quantum number, i.e., phonon number \cite{tuidao}. In ${{\hat{H}}_{\text{ext}}}$, the vibrational mode is driven by an external field with driving strength $F$ and frequency $\omega_{d} $ for investigating synchronization with an external drive.
	
\begin{figure}[ht!]
		\centering\includegraphics[width=10cm]{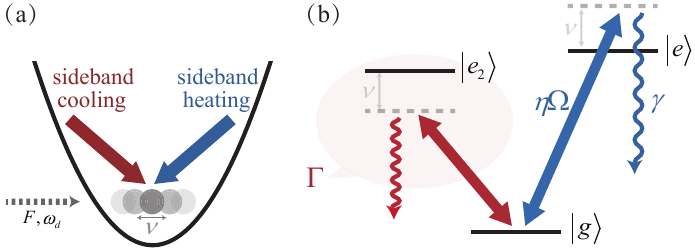}
		\caption{(a) Sketch of the model for trapped-ion system. The ion vibrates in the trap at frequency $\nu$. The ion's vibration can be manipulated simultaneously via the sideband heating and cooling processes, respectively. The heating and cooling compete with each other to maintain a stable amplitude for the phonon laser. Additionally, the vibrational mode of the ion is driven by an external field with driving strength $F$ and frequency $\omega_{d}$. (b) Internal electronic levels of the single ion. The ion undergoes a transition from $\left|g\right\rangle$ to $\left|e\right\rangle$ via the blue sideband resonance, accompanied by adding a phonon. Subsequently, the ion undergoes effective spontaneous emission at a rate $\gamma$. These two processes complete one cycle of sideband heating. The effective damping of the mechanical oscillation at a rate $\Gamma$ is accomplished by the sideband cooling on another different electronic transition $\left|g\right\rangle\leftrightarrow\left|e_2\right\rangle$ and $\left|e_2\right\rangle$ is eliminated adiabatically.}
		\label{fig0}
	\end{figure}

	To generate a phonon laser, in addition to heating the ion's vibrational motion via the blue sideband transition, two dissipation mechanisms have to be added, which is shown in Fig. \ref{fig0}. After the blue sideband transition, the phonon number has increased by one. In order to add more phonons, the single ion has to be reset to the ground state $|g\rangle$ on timescales smaller than the oscillator decay rate, i.e., $\gamma\gg\Gamma$, which can be achieved by effective spontaneous emission of $|e\rangle$ at a rate $\gamma$ through coupling $|e\rangle$ to a short-life level (not shown in Fig. \ref{fig0}), which quickly decays back to $|g\rangle$. Moreover, effective damping of the mechanical oscillation is essential for maintaining a stable amplitude of the phonon laser. This can be achieved by coupling another transition $\left|g\right\rangle\leftrightarrow\left|e_2\right\rangle$ through the red sideband cooling. The realization of these two dissipations and the parameter conditions are described in detail in the supplement. 
	
	Moving to an interaction picture by using a standard time-dependent unitary transformation, $U(t) = \exp\left\{-it\left[\left(\omega_{L}-\omega_{d}\right){\hat{\sigma}_{z}}/2+\omega_{d}\hat{a}^{\dagger}\hat{a}\right]\right\}$, along with the rotating wave approximation, the total Hamiltonian can be transformed into the following form,
	\begin{equation}
		{{\hat{H}}_{\text{I}}}=-\Delta {{\hat{a}}^{\dagger }}\hat{a}+\frac{1}{2}\Delta {{\hat{\sigma }}_{z}}+i\eta \Omega \left({{\hat{a}}^{\dagger }}{{\hat{\sigma }}_{+}}-\hat{a}\hat{\sigma}_{-} \right)+iF\left( {{{\hat{a}}}^{\dagger }}-\hat{a} \right),
		\label{eq4}
	\end{equation}
	where the detuning $\Delta = \omega_{d} -\nu $ is defined as the difference between the external driving frequency $\omega_{d} $ and the ion's vibrational frequency $\nu$. The specific derivation process of Eq. (\ref{eq4}) is provided in the supplement. By considering the dissipation mechanism, the system's dynamics under the Markov approximation \cite{qs2019prl} obeys the following Lindblad master equation, 
	\begin{equation}
		\dot{\hat{\rho }}=-i\left[ {{{\hat{H}}}_{\text{I}}},\hat{\rho } \right]+\gamma \mathcal{D}\left[ {{{\hat{\sigma }}}_{-}} \right]\hat{\rho }+\Gamma\mathcal{D}\left[\hat{a}\right]\hat{\rho },
		\label{eq5}
	\end{equation}
	where $\hat{\rho }$ is the system density matrix and $\mathcal{D}[\hat{O}]\hat{\rho }=\hat{O}\hat{\rho }{{\hat{O}}^{\dagger }}-\left\{ {{{\hat{O}}}^{\dagger }}\hat{O},\hat{\rho } \right\}/2$ is the standard Lindblad dissipator. The numerical simulations for this master equation can be performed using QuTiP \cite{qutip1,qutip2}. Before proceeding further, it is worth noting that the mean-field approximation can be employed to capture the key features of this system. This approximation works well when the number of atoms or phonons is large \cite{pra1994}. In the present model, there is a single ion and the phonon number should be small to be in the quantum regime. Nevertheless, the mean-field approximation still serves as a useful semi-classical method for identifying the appropriate parameter space. By neglecting noise terms and weak quantum correlations between the internal and external degrees of freedom of the ion, the classical mean-field dynamics equations derived from Eq. (\ref{eq5}) via expectation values $\langle {\dot{\hat{O}}} \rangle =\text{Tr}\left\langle \hat{O}\dot{\hat{\rho} } \right\rangle $, read as follows,
	\begin{equation}
		\begin{aligned}
			{{{\dot{x}}}_{1}}&=\eta \Omega{{x}_{2}}-\frac{\Gamma}{2}{{x}_{1}}-\Delta{{y}_{1}}+F,\\
			{{{\dot{y}}}_{1}}&=\eta \Omega{{y}_{2}}-\frac{\Gamma}{2}{{y}_{1}}+\Delta{{x}_{1}},\\
			{{{\dot{x}}}_{2}}&=-\eta\Omega{{x}_{1}}z-\frac{\gamma }{2}{{x}_{2}}-\Delta{{y}_{2}},\\  
			{{{\dot{y}}}_{2}}&=-\eta \Omega{{y}_{1}}z-\frac{\gamma }{2}{{y}_{2}}+\Delta{{x}_{2}},\\ 
			\dot{z}&=4\eta \Omega\left( {{x}_{1}}{{x}_{2}}+{{y}_{1}}{{y}_{2}} \right)-\gamma \left(z+1\right),\\ 
			\label{eq6}
		\end{aligned}
	\end{equation}
	with mean values $\left\langle {\hat{a}} \right\rangle ={{x}_{1}}+i{{y}_{1}},\left\langle {{{\hat{\sigma }}}_{-}} \right\rangle ={{x}_{2}}+i{{y}_{2}},\left\langle {{{\hat{\sigma }}}_{z}} \right\rangle =z$. This set of equations can be utilized to analyze the system's nonlinear dynamics, to find the parameter regions of lasing or synchronization without noise. The derivation of the above equation and the analysis of the stability of the fixed points are given in the supplement.
	
\begin{figure}[ht!]
		\centering\includegraphics[width=10cm]{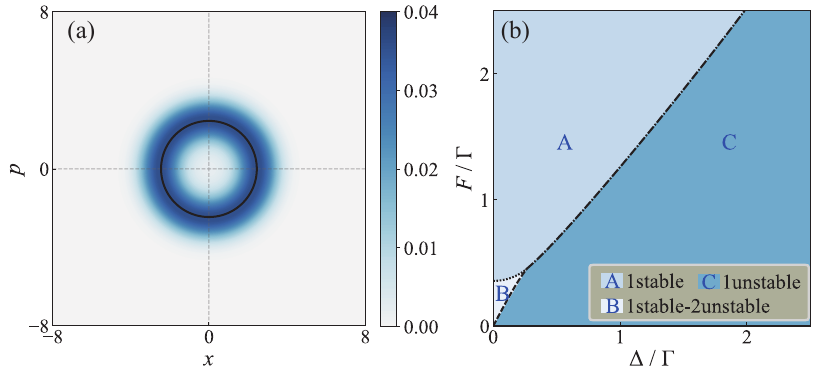}
		\caption{(a) Steady-state Wigner function distribution $W(x,p)$ of an undriven phonon laser ($F/\Gamma =0$). The solid black line represents the limit cycle attractor under the mean-field approximation. (b) Synchronization phase diagram without noise. Regions A, B, and C are divided by the number and stability of fixed points. The dash-dotted line and dashed line indicate the Hopf bifurcation and the saddle-node bifurcation, respectively. They correspond to noiseless boundaries from synchronization to unsynchronization. The fixed parameters for the system are $\eta =0.1,\Omega /\Gamma =25,$ and $\gamma /\Gamma =10$.}
		\label{fig1}
	\end{figure}

	The phonon laser, referred to as a mechanical self-sustained oscillation, is essentially a limit cycle in terms of nonlinear dynamics. In the absence of driving ($F/\Gamma=0$), the threshold for the mechanical self-sustained oscillation is determined by the critical Rabi frequency $\Omega_{\text{th}}=\sqrt{\gamma \Gamma}/(2\eta)$, which signs a phase transition into the lasing phase when $\Omega > \Omega_{\text{th}}$. The amplitude of the limit cycle is given by $\sqrt{{\gamma}/(2\Gamma)-{\gamma^2}/{(8\eta^2\Omega^2)}}$, and it saturates at a large Rabi frequency to be $\sqrt{{\gamma}/{2\Gamma}}$. The ion's mechanical states can be visualized by the Wigner distribution function, defined as 
	$W(x, p)=\int_{-\infty}^{\infty}d se^{ i s p}\langle x+s/2|\hat{\rho}_\mathrm{ph}| x-s/2\rangle /(2\pi)$ 
	\cite{quantumoptics}, where $\hat{\rho}_\mathrm{ph}$ is the reduced density matrix of the ion's vibration after tracing out the internal degrees of freedom, $x$ and $p$ represent the eigenvalues of the operators $\hat{x}=(\hat{a}+\hat{a}^\dagger)/\sqrt{2}$ and $\hat{p}=(\hat{a}-\hat{a}^\dagger)/(\sqrt{2}i)$, respectively. As shown in Fig. \ref{fig1}(a), the steady-state Wigner distribution function of the ion's vibrational mode in the absence of an external drive, exhibits a ring-shaped profile with broadening, indicating limit cycle motion. However, it should be noted that this ring-shaped distribution is only observable under the condition $\gamma \gg \Gamma$, given that the Rabi frequency $\Omega$ cannot be too large to avoid off-resonant transitions in practical experiments. When considering a moderate Rabi frequency and $\gamma \leq \Gamma$, the radius of the limit cycle becomes very small. This, combined with the presence of quantum fluctuations, tends to obscure the ring attractor, making the ring-shaped Wigner distribution disappear. Fig. \ref{fig1}(b) is the corresponding synchronization phase diagram under the mean-field approximation regarding the strength and detuning of the external drive. By neglecting noise effects, a clear boundary between synchronization and unsynchronization regions can be defined. In the frame rotating with the driving frequency, synchronization with the external drive implies the existence of a stable fixed point in the system. Regions A and B both possess a stable fixed point, and their transition to the unsynchronization region C is characterized by two distinct bifurcations. A Hopf bifurcation indicates the onset of unsynchronized behavior for strong driving strength, while a saddle-node bifurcation indicates it for weak ones. This synchronization phase diagram exhibits similarities to that of vdP oscillators, as analyzed in Ref. \cite{qs2014prl}. Both diagrams display three distinct regions and two types of bifurcations. However, it is important to note that the set of nonlinear first-order equations governing the vdP oscillator has a dimension of two, which differs from the five dimensions described by Eq. (\ref{eq6}). This observation suggests that the dissipation mechanism and dimensions do not play a significant role in semi-classical nonlinear dynamics. Future work could aim at proving the prediction that such bifurcation is universal when a limit cycle system is subjected to an external drive.
	
	\section{Steady-state entanglement}
	\label{sec:ent}
	
\begin{figure}[ht!]
		\centering\includegraphics[width=10cm]{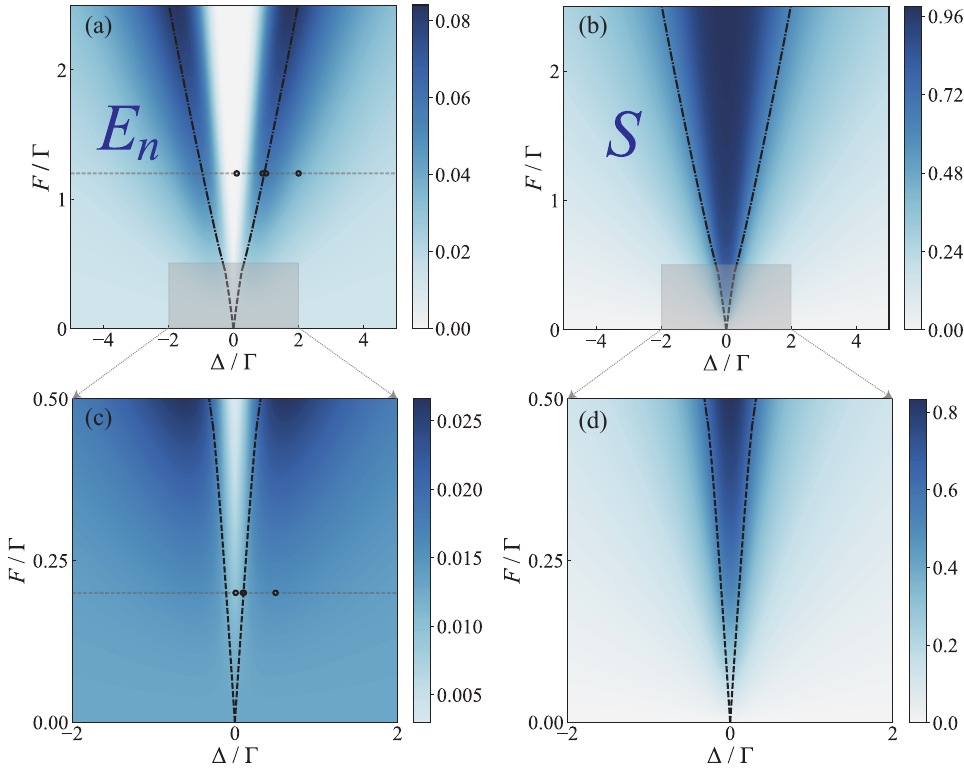}
		\caption{Comparison between (a) the steady-state entanglement ${E}_{n}$ and (b) synchronization measure $S$ in the two-dimensional parameter space with respect to driving detuning $ \Delta / \Gamma$ and driving strength $ F / \Gamma$. The system parameters and dashed or dash-dotted lines denoting boundaries are the same as those in Fig. \ref{fig1}. The hollow black dots in Fig. \ref{fig2}(a) correspond to the parameters $\Delta /\Gamma =0.1,0.9,1.0,2.0$ from left to right in Fig. \ref{fig3}, while the hollow black dots in Fig. \ref{fig2}(c) correspond to the parameters $\Delta /\Gamma =0.01,0.10,0.11,0.50$ from left to right in Fig. \ref{fig4}.}
		\label{fig2}
	\end{figure}

	To investigate the entanglement between the ion's internal and external states, the logarithmic negativity ${{E}_{n}={\log }_{2}{{\left\| {\hat{\rho}}^{\text{T}_\text{A}} \right\|}_{1}}}$ \cite{ent} is used as a measure of entanglement, where $\hat{\rho}^{\text{T}_A}$ is the partially transposed density matrix concerning A, and ${{\left\| \,\cdot \, \right\|}_{1}}$ is the trace norm. In the absence of an external drive, the entanglement between the ion’s internal and external states is relatively low, with a value of only $10^{-2}$. Fig. \ref{fig2}(a) illustrates the steady-state entanglement as a function of the external driving strength $F/\Gamma$ and the frequency detuning $\Delta/\Gamma$. As the driving detuning decreases and approaches resonance, the entanglement drops to zero. This behavior arises because the ion's external states become strongly synchronized with the external drive, resulting in fewer correlations with the internal states. When the detuning is significantly large, the effect of the external drive on the phonon laser is weak, leading to entanglement close to the case without driving. The distribution of entanglement provides clear indications of the synchronization region, with the maximum value of entanglement aligning closely with the noiseless boundary. For low driving strength ($F/\Gamma<0.5$), although the maximum value of entanglement slightly deviates from the boundary, it remains close to zero deep within the Arnold tongue (region B in Fig. \ref{fig1}(b)).
	Due to the inevitable noise in quantum systems, there is no clear boundary between synchronous and non-synchronous regions. However, various measures \cite{qs2013measure,qs2016talita,qs2017measure,qs2020measure} can be employed to quantify quantum synchronization, providing an indication of the relative degree of synchronization. One such measure is defined as $S=\left| S \right|{{e}^{i\theta }}={\left\langle {\hat{a}} \right\rangle }/{\sqrt{\left\langle {{{\hat{a}}}^{\dagger}}\hat{a} \right\rangle }}$ \cite{qs2016talita}, which characterizes the phase locking to a specific value. $\left| S \right|=1$ suggests that the system is perfectly locked, while $\left| S \right|=0$ indicates perfect unlocking. As the driving strength increases, the range of frequency detuning corresponding to localized phase locking widens, resulting in the Arnold tongue depicted by $S$ in Fig. \ref{fig2}(b). Compared to the steady-state entanglement, this measure exhibits a distribution that only shows a monotonic trend of change without any discernible boundaries.
	
	\section{Entanglement dynamics}
	\label{sec:dynamics}
	\subsection{Cross-boundary behavior}
\begin{figure}[ht!]
		\centering\includegraphics[width=10cm]{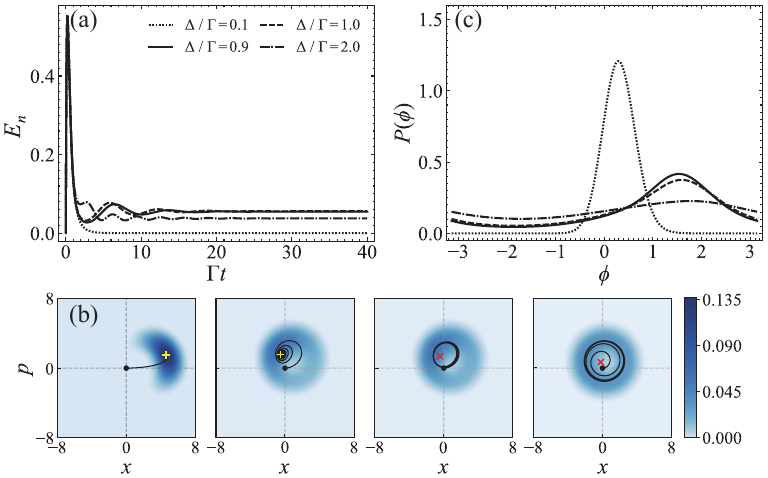}
		\caption{The cross-boundary behaviors at the Hopf bifurcation for a driving strength $ F / \Gamma=1.2$ under various detunings parameters $\Delta /\Gamma =0.1,0.9,1.0,2.0$ (shown in Fig. \ref{fig2}(a)). (a) The entanglement dynamics versus $\Gamma t$ starting from the initial state $\left| g \right\rangle \left| 0 \right\rangle $. (b) Mean-field classical trajectory and corresponding final Wigner function distribution from the left to right with increasing detuning. The initial condition $(x=0,p=0)$ is depicted by a black dot, while the stable and unstable fixed points are represented by yellow plus and red cross signs, respectively. (c) Steady-state phase distribution identical to the line marks in (a). Other parameters are the same as those in Fig. \ref{fig1}.}
		\label{fig3}
	\end{figure}

\begin{figure}[ht!]
		\centering\includegraphics[width=10cm]{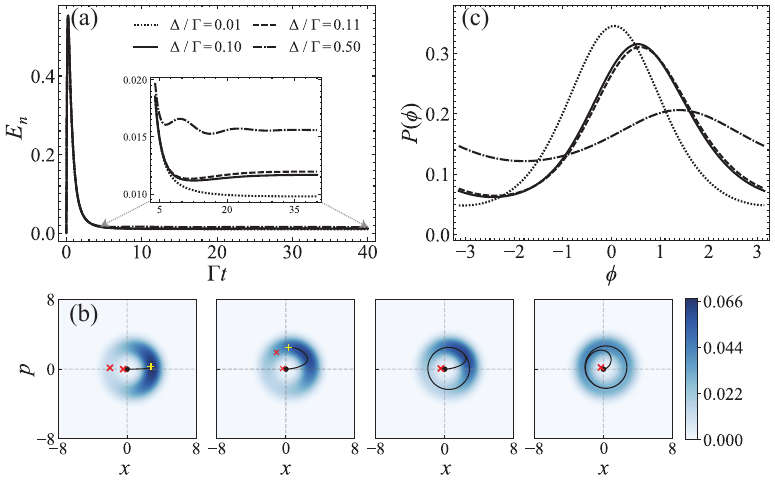}
		\caption{The cross-boundary behaviors at the saddle-node bifurcation for a driving strength $ F / \Gamma=0.2$ under various detuning parameters $\Delta /\Gamma =0.01,0.10,0.11,0.50$ (shown in Fig. \ref{fig2}(c)). The contents of the figure and all the other parameters are identical to Fig. \ref{fig3}.} 
		\label{fig4}
	\end{figure}

	The characteristics of steady-state entanglement at the noiseless boundary inspire us to investigate the evolution of entanglement. Figs. \ref{fig3}(a) and \ref{fig4}(a) show the entanglement dynamics for driving strength ${F/ \Gamma=1.2}$ and ${F/\Gamma=0.2}$, respectively, starting from the initial state $\left| g \right\rangle \left| 0 \right\rangle $, which is the product state of the internal ground state and the external vacuum state. For almost all parameters, entanglement rapidly reaches a peak, suggesting that it follows a transient process. This behavior can be attributed to the immediate coupling between the internal and external degrees of freedom of the ion under the blue sideband process. Noticeable oscillation behaviors in entanglement can be observed for parameters with a large steady-state entanglement, and these oscillations maintain relatively high values during the decay towards the steady state. Figs. \ref{fig3}(a) and \ref{fig4}(a) also show that the frequency of the oscillations increases with the detuning. Additionally, when the detuning is small, the entanglement does not exhibit any oscillation, but for large detuning, it oscillates with small amplitudes.
	
	The oscillatory behavior can be partially captured by the classical trajectories obtained through the mean-field approximation. In Figs. \ref{fig3}(b) and \ref{fig4}(b), the classical trajectories starting from the origin, as well as the distribution of the Wigner function of the final evolved state are depicted. In the early stages of evolution, the quantum quasi-distribution closely follows the classical trajectory. The oscillation dynamics are related to the spiral rotation towards a stable fixed point or following a limit cycle attractor. Although cross-boundary behavior can manifest through two distinct types of bifurcations in terms of the classical trajectory, both transitions involve a shift from falling towards a stable fixed point to continuous orbiting along a limit cycle. For relatively strong driving strength, a Hopf bifurcation occurs once the noiseless boundary is crossed. The stable fixed point initially moves counterclockwise, loses stability, and eventually moves towards the origin. The Wigner function distribution gradually spreads out along a ring shape, forming either a localized blob or a broadened circular distribution. With weak driving strength, the saddle-node bifurcation involves the movement of the stable fixed point along a cycle in response to increasing detuning until it merges and disappears with another unstable fixed point at the bifurcation boundary, leaving only an unstable fixed point close to the origin. The change in the Wigner function distribution is less pronounced compared to the case of strong driving. It is evident that the transition from synchronous to asynchronous regions is not well-defined. With an increase in detuning, the phase distribution $P\left( \phi \right)=\sum\nolimits_{n,m=0}^{\infty }{{{e}^{i(m-n)\phi }}}\left\langle n\left| {\hat{\rho}_\mathrm{ph}} \right| m \right\rangle /2\pi$ tends smoothly towards flatness (see Figs. \ref{fig3}(c) and \ref{fig4}(c)). The presence of quantum noise results in a blurring of the transition, leading to no discernible difference in the phase distribution near the noiseless boundary in comparison to other parameter regions. In general, only entanglement displays distinctive dynamical characteristics near the noiseless synchronization boundary.
	
	\subsection{Properites of Liouvillian eigenspectrum}
	The dynamics of entanglement can be better understood by examining the Liouvillian eigenspectrum. In the form of the Liouvillian superoperator $\mathcal{L}$, the master equation is given by ${{\partial }_{t}}\hat{\rho }=\mathcal{L}\hat{\rho }$. The dynamics of system can be further analyzed by the eigenequation, i.e., $\mathcal{L}{{\hat{\rho }}_{i}}={{\lambda }_{i}}{{\hat{\rho }}_{i}}$, where the eigenvalues form the Liouvillian eigenspectrum. Moreover, as $\mathcal{L}$ is not Hermitian, it exhibits the property of having left eigenmatrices ${{\mathcal{L}}^{\dagger }}{{\hat{\sigma }}_{j}}=\lambda _{j}^{*}{{\hat{\sigma }}_{j}}$ and satisfies $\operatorname{Tr}\left[ {{{\hat{\rho }}}_{i}}{{{\hat{\sigma }}}_{j}} \right]={{\delta }_{ij}}$ \cite{lep2019two,lep2019,lep2020pra,lep2020nori}. For each eigenmode, the decay rate and frequency can be defined by the real and imaginary parts of the corresponding eigenvalue, ${{\Gamma}_{j}}=\left| \operatorname{Re}\left[ {{\lambda }_{j}} \right] \right|$ and ${{\nu}_{j}}=\text{Im}\left[ {{\lambda }_{j}} \right]$, respectively. The real parts of these eigenvalues are non-positive, and it is convenient to sort the eigenvalues and eigenmatrices in such a way that $\Gamma_0\le\Gamma_1\le\Gamma_2\le\dots$. If $\mathcal{L}$ is diagonalizable, the density matrix of the system can be represented as \cite{qs2021jiadai},
	\begin{equation}
		\hat{\rho }(t)={{\hat{\rho }}_{ss}}+\underset{j\ge1}{\mathop{\sum }}\,\operatorname{Tr}\left[ {{{\hat{\sigma }}}_{j}}\hat{\rho }(0) \right]{{\hat{\rho }}_{j}}{{e}^{{{\lambda }_{j}}t}}, 
	\end{equation}
	where $\hat{\rho}_{ss}= {\hat{\rho }_{0}} / \text{Tr} \left[ \hat{\rho }_{0} \right ] $ is the steady state corresponding to the eigenvalue $\lambda_{0}=0$. 
	
\begin{figure}[ht!]
		\centering\includegraphics[width=12cm]{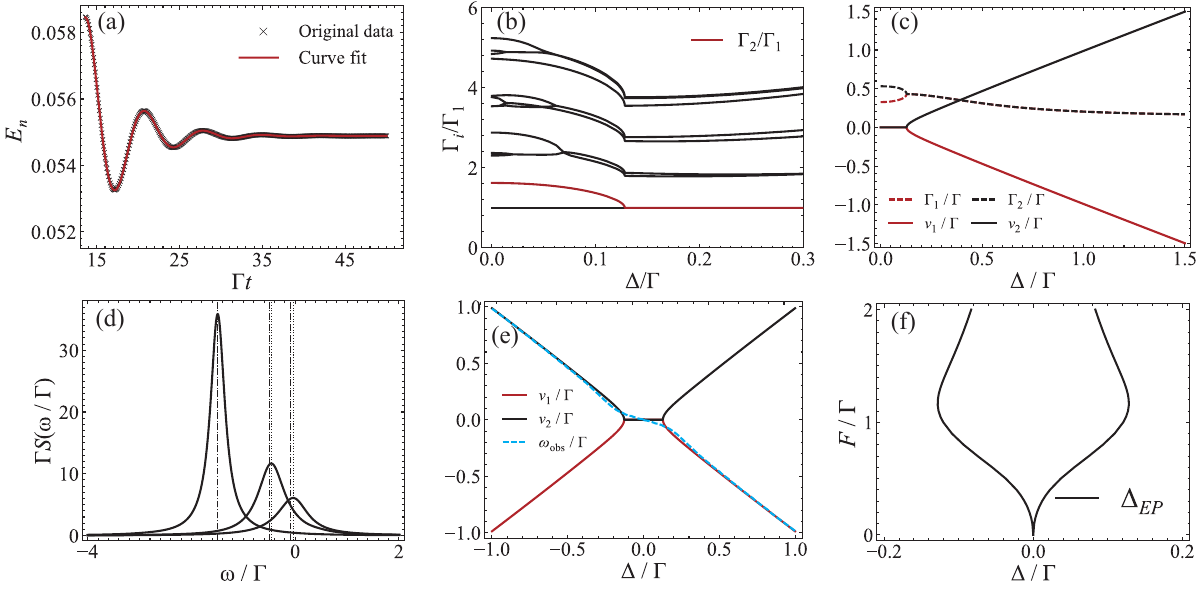}
		\caption{Liouvillian eigenspectrum and Frequency entrainment. (a) Fitting of entanglement dynamics $(F/\Gamma=1.2,\Delta/\Gamma=0.9)$ considering only $t\gg\tau={{\Gamma}_{3}^{-1}}$, with $\Gamma_1/\Gamma=0.217, \nu_1 /\Gamma=-0.885$. The fitted expression is $0.073\exp{(-0.219\Gamma t)}\sin{(-0.886\Gamma t}+1.323)+0.055$ and the fitting parameters for the oscillatory decay curve are as follows: the amplitude is 0.073, the decay coefficient is 0.219, the oscillation frequency is $-0.886$, the initial phase is 1.323, and the vertical offset is 0.055. (b) The ratio of decay rates $\Gamma_{i}/\Gamma_{1}$ for the first few eigenvalues and (c) decay rates and frequencies of the first two eigenmodes for $F/\Gamma=1.2$. (d) Power spectrum $S(\omega)$ for $F/ \Gamma =1.2$ with detuning $\Delta /\Gamma =1.5,0.5,0.1$ from left to right. The dotted vertical lines represent ${{\omega }_{\text{obs}}}$ while the dash-dotted lines represent the intrinsic frequency $-\Delta $ of the vibrational mode in the rotating frame. (e) Comparison between ${{\omega }_{\text{obs}}}$ and $\nu_{1,2}$. (f) Similar Arnold tongue formed by LEPs.}
		\label{fig5}
	\end{figure}

	In Fig. \ref{fig5}(a), the dynamics of entanglement is fitted by an exponentially decaying oscillation. It is observed that the decay rate and oscillation frequency of the entanglement align well with the values of $\Gamma_{1}$ and $\nu_{1}$, respectively. The reason for that is the later evolution of the system $(t\gg\tau={{\Gamma}_{3}^{-1}})$ is primarily governed by the first two eigenmodes. As shown in Fig. \ref{fig5}(b), the decay rates of other modes with $i>2$ are several times as much as $\Gamma_{1}$ and decay rapidly, only impacting the transient dynamics \cite{lep2016,qs2021jiadai}. It is evident from this subplot that all of the $\Gamma_{i}/\Gamma_{1}$ experience a turning point at the same location. Building upon this observation, the decay rates and eigenfrequencies of the first two eigenmodes with respect to $\Delta / \Gamma$ are given in Fig. \ref{fig5}(c), which shows the appearance of LEPs ($\Delta ={{\Delta }_{\text{EP}}}$). At LEPs, ${{\lambda }_{1,2}}$ becomes equal, and the corresponding eigenmatrices merge together. When the detuning satisfies $\Delta <{{\Delta }_{\text{EP}}}$, ${{\lambda }_{1,2}}$ become real-valued, thus explaining the absence of oscillation for the entanglement evolution in this region. For $\Delta >{{\Delta }_{\text{EP}}}$, ${{\lambda }_{1,2}}$ are complex conjugates with ${{\Gamma}_{1}}={{\Gamma}_{2}}$ and the entanglement oscillates at a frequency $\left | \nu_{1} \right |=\left | \nu_{2} \right |$, which equals $\Delta $ when the detuning is slightly far away from ${{\Delta }_{\text{EP}}}$. This frequency distribution is similar to the phenomenon of frequency entrainment in synchronization without noise \cite{nonlinear}.
	
	Frequency entrainment is one of the main features of forced synchronization, i.e., the frequency of limit cycle motion is adjusted to that of the external drive. It can be discussed in the context of quantum synchronization via the power spectrum \cite{qs2014prl}, which is defined as $S(\omega )=\int_{-\infty }^{\infty }{{{\left\langle {{{\hat{a}}}^{\dagger }}(t)\hat{a}(0) \right\rangle }_{ss}}}{{e}^{-i\omega t }}dt$, where ${{\left\langle {{{\hat{a}}}^{\dagger }}(t)\hat{a}(0) \right\rangle }_{ss}}$ denotes the amplitude two-time correlation in the stationary state. The observed frequency ${{\omega }_\text{obs}}$ can be determined by its maximum, i.e., ${{\omega }_\text{obs}}=\text{argmax}(S(\omega ))$. This maximum corresponds to the central frequency of the phonon laser in the rotating frame. In Fig. \ref{fig5}(d), the power spectrum $S(\omega)$ is shown for a fixed external driving strength with different detunings. If the spectrum exhibits a single peak at $\omega =0$, it indicates perfect synchronization between the phonon laser and the external drive. However, in the presence of quantum noise, perfect synchronization is not achievable. The observed frequency ${{\omega }_\text{obs}}$ is pulled towards the driving frequency but does not reach it. The difference between the observed frequency and the intrinsic frequency decreases as the driving detuning increases, indicating a gradual weakening of the synchronization effect.
	
	The comparison between the observed frequency ${{\omega }_{\text{obs}}}$ and the imaginary parts of the first two eigenvalues ${{\nu}_{1,2}}$ is shown in Fig. \ref{fig5}(e). It can be seen that the curves of the observed frequency follow the eigenfrequencies very well, except for deviating a little bit around the LEPs. This can be understood from the structure of the amplitude two-time correlation ${{\left\langle {{{\hat{a}}}^{\dagger }}(t )\hat{a}(0) \right\rangle }_{ss}}$ mentioned above in terms of the Liouvillian eigenmodes \cite{qs2021jiadai},
	\begin{equation}
		{{\left\langle {{{\hat{a}}}^{\dagger }}(t )\hat{a}(0) \right\rangle }_{ss}}=\underset{j\ge1}{\mathop{\sum }}\,\operatorname{Tr}\left[ \hat{\sigma }_{j}\hat{a}{{{\hat{\rho }}}_{ss}} \right]\operatorname{Tr}\left[ {{{\hat{a}}}^{\dagger }}{{{\hat{\rho }}}_{j}} \right]{{e}^{{{\lambda }_{j}} t }}.
	\end{equation}
	The power spectrum $S(\omega)$ is obtained by taking the Fourier transform of the correlation function, which can be expressed as a sum of Lorentzian profiles centered at the imaginary parts of the eigenvalues of $\mathcal{L}$ while the width is given by the real parts and weighted by some coefficients \cite{lep2022}. In the case where $\nu_{1,2}$ does not align with $\omega_\text{obs}$, the contribution of the adjacent eigenvalues should be taken into account in addition to that of the first two eigenvalues. When the detuning is relatively small, the external drive can effectively draw the phonon frequency closer to its own frequency, leading to a plateau in the range of $\nu_{1,2}=0$ and $\omega_\text{obs}$ approaches but does not completely overlap with this plateau due to the presence of noise. Anyway, the behaviors of the first two eigenvalues play a major role, and the appearance of LEPs can be considered as an important indicator of frequency entrainment in quantum synchronization. Furthermore, the distribution of LEPs is shown in an attempt to construct the Arnold tongue in Fig. \ref{fig5}(f). The value of ${{\Delta }_{\text{EP}}}$ shows an upward trend with increasing driving strength $(F/ \Gamma<1.15)$, indicating that a larger range of synchronization can occur with a stronger driving strength. However, as the driving strength continues to increase $(F/ \Gamma >1.15)$, ${{\Delta }_{\text{EP}}}$ decreases due to the influence of the stronger driving strength on the amplitude of the limit cycle, rather than solely modifying its phase. Moreover, the locations of LEPs are in the deep synchronization region, where the entanglement is quite small. This is reasonable since the noise will narrow the frequency range of synchronization compared with the noiseless boundary.
	
	\section{Conclusion}
	\label{sec:conclusion}
	To conclude, entanglement signatures for quantum synchronization with a single-ion phonon laser to an external drive have been investigated from the perspectives of both steady-state distribution and dynamical evolution. In the presence of quantum noise, various measures of quantum synchronization such as phase distribution do not exhibit distinctive features near the mean-field noiseless boundary from synchronization to unsynchronization. However, the steady-state entanglement between the internal and external states of a single ion reaches its maximum near the noiseless boundary, with corresponding evolution displaying noticeable oscillatory behavior, thus providing a good indicator for quantum synchronization. 
	
	Current research on the relationship between quantum correlations and quantum synchronization primarily focuses on the mutual synchronization of multiple limit cycle systems \cite{qs2012,qs2013measure,qs20131, qs2014tongue,qs2015huxinxi, qs2016, qs2018,qs2019gl}. In this case, both synchronization and quantum correlations are due to interactions between two or more limit cycle systems, and it is therefore natural to investigate the quantum signatures of synchronization. Most of the previous works show that quantum correlations such as mutual information, and quantum entanglement can produce the same distribution pattern in parameter space as the defined synchronization measures, indicating the stronger the synchronization, the larger the quantum correlations \cite{qs2014tongue,qs2015huxinxi,qs2018,qs2019gl}. In this work, the situation of forced synchronization is explored, where there is only one limit cycle system subjected to an external drive. To create the limit cycle motion of the ion's vibration, the internal levels are utilized to have nonlinear interactions with the vibration. Due to the interaction, there exists entanglement between the ion's internal and external degrees of freedom, thus allowing for the exploration of the relationship between entanglement and synchronization. However, the entanglement distribution pattern is different from that of the synchronization measure, with the remarkable new feature that the maximum entanglement is near the noiseless boundary from synchronization to unsynchronization. This work extends the study on the interplay of entanglement and synchronization from mutual case to forced case, from entanglement between different systems to that of different degrees of freedom of a single system. However, if the vibration itself is nonlinear and no auxiliary degrees of freedom are involved, then the discussion of entanglement is not possible. 
	
	Furthermore, frequency entrainment can be inferred from the later time evolution of entanglement, as indicated by the favorable consistency between the observed frequency and the imaginary components of the first two eigenvalues in the Liouvillian eigenspectrum. In the context of quantum synchronization, the frequency entrainment is manifested as the frequency of limit cycle motion being pulled towards that of the external drive, resulting in a plateau region in the diagram of observed frequency as a function of external driving detuning. The appearance of LEPs in the first two eigenvalues also shows a plateau with respect to the driving detuning. These two phenomena match so well that the existence of LEPs is a strong indication of frequency entrainment and should be observed in similar models of forced synchronization. Around the LEPs, chiral state transfer \cite{lep2022qubit} and the enhancement of quantum heat engine efficiency \cite{lep2023he} have been studied, which offers the possibility of combining quantum synchronization with the related applications.

	\begin{backmatter}
		\bmsection{Funding}
		National Natural Science Foundation of China (Grant Nos. 11574398, 12174448, 12174447, 11904402, 12074433, 11871472, 12004430 and 12365006).
		\bmsection{Acknowledgments}
		Z. J. Deng is grateful to Jia-Hao Huang, Shuo Zhang, and Yan-Li Zhou for useful discussions. 
		\bmsection{Disclosures}
		The authors declare no conflicts of interest.
		
		\bmsection{Data Availability Statement}
		Data underlying the results presented in this paper are not publicly available at this time but may
		be obtained from the authors upon reasonable request.
		
		\bmsection{Supplemental document}
		See Supplement 1 for supporting content.
		
	\end{backmatter}
	
	\bibliography{sample}

\begin{thebibliography}{10}
\newcommand{\enquote}[1]{``#1''}

\bibitem{nonlinear}
A.~Pikovsky, M.~Rosenblum, and J.~Kurths, \emph{{Synchronization: A Universal
  Concept in Nonlinear Sciences}}, Cambridge Nonlinear Science Series
  (Cambridge University Press, 2001).

\bibitem{qs20180}
A.~Roulet and C.~Bruder, \enquote{Synchronizing the smallest possible system,}
  {\protect\JournalTitle{Phys. Rev. Lett.}} \textbf{121}, 053601 (2018).

\bibitem{qs2018}
A.~Roulet and C.~Bruder, \enquote{Quantum synchronization and entanglement
  generation,} {\protect\JournalTitle{Phys. Rev. Lett.}} \textbf{121}, 063601
  (2018).

\bibitem{qs2020spin1}
A.~W. Laskar, P.~Adhikary, S.~Mondal, P.~Katiyar, S.~Vinjanampathy, and
  S.~Ghosh, \enquote{{Observation of Quantum Phase Synchronization in Spin-1
  Atoms},} {\protect\JournalTitle{Phys. Rev. Lett.}} \textbf{125}, 013601
  (2020).

\bibitem{qs2023}
L.~Zhang, Z.~Wang, Y.~Wang, J.~Zhang, Z.~Wu, J.~Jie, and Y.~Lu,
  \enquote{Quantum synchronization of a single trapped-ion qubit,}
  {\protect\JournalTitle{Phys. Rev. Res.}} \textbf{5}, 033209 (2023).

\bibitem{qs2013prl}
T.~E. Lee and H.~R. Sadeghpour, \enquote{{Quantum Synchronization of Quantum
  van der Pol Oscillators with Trapped Ions},} {\protect\JournalTitle{Phys.
  Rev. Lett.}} \textbf{111}, 234101 (2013).

\bibitem{qs2014prl}
S.~Walter, A.~Nunnenkamp, and C.~Bruder, \enquote{{Quantum Synchronization of a
  Driven Self-Sustained Oscillator},} {\protect\JournalTitle{Phys. Rev. Lett.}}
  \textbf{112}, 094102 (2014).

\bibitem{qs2014tongue}
T.~E. Lee, C.-K. Chan, and S.~Wang, \enquote{Entanglement tongue and quantum
  synchronization of disordered oscillators,} {\protect\JournalTitle{Phys. Rev.
  E}} \textbf{89}, 022913 (2014).

\bibitem{qs2017pra}
T.~Weiss, S.~Walter, and F.~Marquardt, \enquote{Quantum-coherent phase
  oscillations in synchronization,} {\protect\JournalTitle{Phys. Rev. A}}
  \textbf{95}, 041802 (2017).

\bibitem{qs2018yasuo}
S.~Sonar, M.~Hajdu\ifmmode~\check{s}\else \v{s}\fi{}ek, M.~Mukherjee, R.~Fazio,
  V.~Vedral, S.~Vinjanampathy, and L.-C. Kwek, \enquote{{Squeezing Enhances
  Quantum Synchronization},} {\protect\JournalTitle{Phys. Rev. Lett.}}
  \textbf{120}, 163601 (2018).

\bibitem{qs2021coherence}
Y.~Kato and H.~Nakao, \enquote{Quantum coherence resonance,}
  {\protect\JournalTitle{New Journal of Physics}} \textbf{23}, 043018 (2021).

\bibitem{qs2021jiadai}
A.~Cabot, G.~L. Giorgi, and R.~Zambrini, \enquote{Metastable quantum
  entrainment,} {\protect\JournalTitle{New Journal of Physics}} \textbf{23},
  103017 (2021).

\bibitem{qs2014atom}
M.~Xu, D.~A. Tieri, E.~C. Fine, J.~K. Thompson, and M.~J. Holland,
  \enquote{{Synchronization of Two Ensembles of Atoms},}
  {\protect\JournalTitle{Phys. Rev. Lett.}} \textbf{113}, 154101 (2014).

\bibitem{qs2016pra}
A.~Roth and K.~Hammerer, \enquote{{Synchronization of active atomic clocks via
  quantum and classical channels},} {\protect\JournalTitle{Phys. Rev. A}}
  \textbf{94}, 043841 (2016).

\bibitem{qs2015hush}
M.~R. Hush, W.-b. Li, S.~Genway, I.~Lesanovsky, and A.~D. Armour, \enquote{Spin
  correlations as a probe of quantum synchronization in trapped-ion phonon
  lasers,} {\protect\JournalTitle{Phys. Rev. A}} \textbf{91}, 061401 (2015).

\bibitem{qs2015wigner}
K.~Shlomi, D.~Yuvaraj, I.~Baskin, O.~Suchoi, R.~Winik, and E.~Buks,
  \enquote{Synchronization in an optomechanical cavity,}
  {\protect\JournalTitle{Phys. Rev. E}} \textbf{91}, 032910 (2015).

\bibitem{qs2016talita}
T.~Weiss, A.~Kronwald, and F.~Marquardt, \enquote{Noise-induced transitions in
  optomechanical synchronization,} {\protect\JournalTitle{New Journal of
  Physics}} \textbf{18}, 013043 (2016).

\bibitem{qs2017gl}
E.~Amitai, N.~L\"orch, A.~Nunnenkamp, S.~Walter, and C.~Bruder,
  \enquote{Synchronization of an optomechanical system to an external drive,}
  {\protect\JournalTitle{Phys. Rev. A}} \textbf{95}, 053858 (2017).

\bibitem{trappedion}
D.~Leibfried, R.~Blatt, C.~Monroe, and D.~Wineland, \enquote{Quantum dynamics
  of single trapped ions,} {\protect\JournalTitle{Rev. Mod. Phys.}}
  \textbf{75}, 281--324 (2003).

\bibitem{pl2009}
K.~Vahala, M.~Herrmann, S.~Knünz, V.~Batteiger, G.~Saathoff, T.~W. Hänsch,
  and T.~Udem, \enquote{A phonon laser,} {\protect\JournalTitle{Nature
  Physics}} \textbf{5}, 682--686 (2009).

\bibitem{pl2013}
Y.~Xie, W.~Wan, H.~Y. Wu, F.~Zhou, L.~Chen, and M.~Feng, \enquote{Pulsed phonon
  lasing in trapped ions,} {\protect\JournalTitle{Phys. Rev. A}} \textbf{87},
  053402 (2013).

\bibitem{pl2010}
S.~Kn\"unz, M.~Herrmann, V.~Batteiger, G.~Saathoff, T.~W. H\"ansch, K.~Vahala,
  and T.~Udem, \enquote{{Injection Locking of a Trapped-Ion Phonon Laser},}
  {\protect\JournalTitle{Phys. Rev. Lett.}} \textbf{105}, 013004 (2010).

\bibitem{pl2021}
Z.~Liu, Y.~Wei, L.~Chen, J.~Li, S.~Dai, F.~Zhou, and M.~Feng,
  \enquote{{Phonon-Laser Ultrasensitive Force Sensor},}
  {\protect\JournalTitle{Phys. Rev. Appl.}} \textbf{16}, 044007 (2021).

\bibitem{forced}
A.~Balanov, N.~Janson, D.~Postnov, and O.~Sosnovtseva, \emph{{Synchronization:
  From Simple to Complex}} (Springer, 2009).

\bibitem{qs2015huxinxi}
V.~Ameri, M.~Eghbali-Arani, A.~Mari, A.~Farace, F.~Kheirandish, V.~Giovannetti,
  and R.~Fazio, \enquote{Mutual information as an order parameter for quantum
  synchronization,} {\protect\JournalTitle{Phys. Rev. A}} \textbf{91}, 012301
  (2015).

\bibitem{qs2021feedback}
Y.~Kato and H.~Nakao, \enquote{Enhancement of quantum synchronization via
  continuous measurement and feedback control,} {\protect\JournalTitle{New
  Journal of Physics}} \textbf{23}, 013007 (2021).

\bibitem{atomlaser}
Y.~Mu and C.~M. Savage, \enquote{One-atom lasers,} {\protect\JournalTitle{Phys.
  Rev. A}} \textbf{46}, 5944--5954 (1992).

\bibitem{ionlaser}
F.~Dubin, C.~Russo, H.~G. Barros, A.~Stute, C.~Becher, P.~O. Schmidt, and
  R.~Blatt, \enquote{Quantum to classical transition in a single-ion laser,}
  {\protect\JournalTitle{Nature Physics}} \textbf{6}, 350--353 (2010).

\bibitem{pl2023}
T.~Behrle, T.~L. Nguyen, F.~Reiter, D.~Baur, B.~de~Neeve, M.~Stadler,
  M.~Marinelli, F.~Lancellotti, S.~F. Yelin, and J.~P. Home, \enquote{{Phonon
  Laser in the Quantum Regime},} {\protect\JournalTitle{Phys. Rev. Lett.}}
  \textbf{131}, 043605 (2023).

\bibitem{gl2014}
M.~Aspelmeyer, T.~J. Kippenberg, and F.~Marquardt, \enquote{Cavity
  optomechanics,} {\protect\JournalTitle{Rev. Mod. Phys.}} \textbf{86},
  1391--1452 (2014).

\bibitem{lep2019two}
N.~Hatano, \enquote{{Exceptional points of the Lindblad operator of a two-level
  system},} {\protect\JournalTitle{Molecular Physics}} \textbf{117}, 2121--2127
  (2019).

\bibitem{lep2019}
F.~Minganti, A.~Miranowicz, R.~W. Chhajlany, and F.~Nori, \enquote{{Quantum
  exceptional points of non-Hermitian Hamiltonians and Liouvillians: The
  effects of quantum jumps},} {\protect\JournalTitle{Phys. Rev. A}}
  \textbf{100}, 062131 (2019).

\bibitem{lep2020pra}
F.~Minganti, A.~Miranowicz, R.~W. Chhajlany, I.~I. Arkhipov, and F.~Nori,
  \enquote{{Hybrid-Liouvillian formalism connecting exceptional points of
  non-Hermitian Hamiltonians and Liouvillians via postselection of quantum
  trajectories},} {\protect\JournalTitle{Phys. Rev. A}} \textbf{101}, 062112
  (2020).

\bibitem{lep2020nori}
I.~I. Arkhipov, A.~Miranowicz, F.~Minganti, and F.~Nori, \enquote{{Liouvillian
  exceptional points of any order in dissipative linear bosonic systems:
  Coherence functions and switching between $\mathcal{PT}$ and
  anti-$\mathcal{PT}$ symmetries},} {\protect\JournalTitle{Phys. Rev. A}}
  \textbf{102}, 033715 (2020).

\bibitem{tuidao}
M.~ŠAŠURA and V.~BUŽEK, \enquote{Cold trapped ions as quantum information
  processors,} {\protect\JournalTitle{Journal of Modern Optics}} \textbf{49},
  1593--1647 (2002).

\bibitem{exp1998}
D.~Wineland, C.~Monroe, W.~Itano, D.~Leibfried, B.~King, and D.~Meekhof,
  \enquote{{Experimental Issues in Coherent Quantum-State Manipulation of
  Trapped Atomic Ions},} {\protect\JournalTitle{J Res Natl Inst Stand Technol}}
  \textbf{103}, 259--328 (1998).

\bibitem{qs2019prl}
S.~Dutta and N.~R. Cooper, \enquote{{Critical Response of a Quantum van der Pol
  Oscillator},} {\protect\JournalTitle{Phys. Rev. Lett.}} \textbf{123}, 250401
  (2019).

\bibitem{qutip1}
J.~Johansson, P.~Nation, and F.~Nori, \enquote{{QuTiP: An open-source Python
  framework for the dynamics of open quantum systems},}
  {\protect\JournalTitle{Computer Physics Communications}} \textbf{183},
  1760--1772 (2012).

\bibitem{qutip2}
J.~Johansson, P.~Nation, and F.~Nori, \enquote{{QuTiP 2: A Python framework for
  the dynamics of open quantum systems},} {\protect\JournalTitle{Computer
  Physics Communications}} \textbf{184}, 1234--1240 (2013).

\bibitem{pra1994}
P.~R. Rice and H.~J. Carmichael, \enquote{{Photon statistics of a cavity-QED
  laser: A comment on the laser-phase-transition analogy},}
  {\protect\JournalTitle{Phys. Rev. A}} \textbf{50}, 4318--4329 (1994).

\bibitem{quantumoptics}
D.~Walls and G.~J. Milburn, eds., \emph{Quantum Optics} (Springer Berlin,
  2008).

\bibitem{ent}
G.~Vidal and R.~F. Werner, \enquote{Computable measure of entanglement,}
  {\protect\JournalTitle{Phys. Rev. A}} \textbf{65}, 032314 (2002).

\bibitem{qs2013measure}
A.~Mari, A.~Farace, N.~Didier, V.~Giovannetti, and R.~Fazio, \enquote{{Measures
  of Quantum Synchronization in Continuous Variable Systems},}
  {\protect\JournalTitle{Phys. Rev. Lett.}} \textbf{111}, 103605 (2013).

\bibitem{qs2017measure}
W.~Li, W.~Zhang, C.~Li, and H.~Song, \enquote{Properties and relative measure
  for quantifying quantum synchronization,} {\protect\JournalTitle{Phys. Rev.
  E}} \textbf{96}, 012211 (2017).

\bibitem{qs2020measure}
N.~Jaseem, M.~Hajdu\ifmmode~\check{s}\else \v{s}\fi{}ek, P.~Solanki, L.-C.
  Kwek, R.~Fazio, and S.~Vinjanampathy, \enquote{Generalized measure of quantum
  synchronization,} {\protect\JournalTitle{Phys. Rev. Res.}} \textbf{2}, 043287
  (2020).

\bibitem{lep2016}
K.~Macieszczak, M.~Gu\c{t}\u{a}, I.~Lesanovsky, and J.~P. Garrahan,
  \enquote{{Towards a Theory of Metastability in Open Quantum Dynamics},}
  {\protect\JournalTitle{Phys. Rev. Lett.}} \textbf{116}, 240404 (2016).

\bibitem{lep2022}
Y.-L. Zhou, D.~Yan, and W.~Li, \enquote{Rydberg electromagnetically induced
  transparency and absorption of strontium triplet states in a weak microwave
  field,} {\protect\JournalTitle{Phys. Rev. A}} \textbf{105}, 053714 (2022).

\bibitem{qs2012}
G.~L. Giorgi, F.~Galve, G.~Manzano, P.~Colet, and R.~Zambrini, \enquote{Quantum
  correlations and mutual synchronization,} {\protect\JournalTitle{Phys. Rev.
  A}} \textbf{85}, 052101 (2012).

\bibitem{qs20131}
G.~L. Giorgi, F.~Plastina, G.~Francica, and R.~Zambrini, \enquote{Spontaneous
  synchronization and quantum correlation dynamics of open spin systems,}
  {\protect\JournalTitle{Phys. Rev. A}} \textbf{88}, 042115 (2013).

\bibitem{qs2016}
C.~Davis-Tilley and A.~D. Armour, \enquote{Synchronization of micromasers,}
  {\protect\JournalTitle{Phys. Rev. A}} \textbf{94}, 063819 (2016).

\bibitem{qs2019gl}
C.-G. Liao, R.-X. Chen, H.~Xie, M.-Y. He, and X.-M. Lin, \enquote{Quantum
  synchronization and correlations of two mechanical resonators in a
  dissipative optomechanical system,} {\protect\JournalTitle{Phys. Rev. A}}
  \textbf{99}, 033818 (2019).

\bibitem{lep2022qubit}
W.~Chen, M.~Abbasi, B.~Ha, S.~Erdamar, Y.~N. Joglekar, and K.~W. Murch,
  \enquote{{Decoherence-Induced Exceptional Points in a Dissipative
  Superconducting Qubit},} {\protect\JournalTitle{Phys. Rev. Lett.}}
  \textbf{128}, 110402 (2022).

\bibitem{lep2023he}
J.-T. Bu, J.-Q. Zhang, G.-Y. Ding, J.-C. Li, J.-W. Zhang, B.~Wang, W.-Q. Ding,
  W.-F. Yuan, L.~Chen, i.~m. c.~K. \"Ozdemir, F.~Zhou, H.~Jing, and M.~Feng,
  \enquote{{Enhancement of Quantum Heat Engine by Encircling a Liouvillian
  Exceptional Point},} {\protect\JournalTitle{Phys. Rev. Lett.}} \textbf{130},
  110402 (2023).

\end{thebibliography}


\begin{thebibliography}{10}
\newcommand{\enquote}[1]{``#1''}

\bibitem{tuidao}
M.~\v{S}A\v{S}URA and V.~BU\v{Z}EK, \enquote{Cold trapped ions as quantum
  information processors,} {\protect\JournalTitle{Journal of Modern Optics}}
  \textbf{49}, 1593--1647 (2002).

\bibitem{laser}
M.~Kardar, \emph{{Statistical Physics of Particles}} (Cambridge University
  Press, 2007).

\bibitem{3981}
D.~Wineland, C.~Monroe, W.~Itano, D.~Leibfried, B.~King, and D.~Meekhof,
  \enquote{{Experimental Issues in Coherent Quantum-State Manipulation of
  Trapped Atomic Ions},} {\protect\JournalTitle{J Res Natl Inst Stand Technol}}
  \textbf{103}, 259--328 (1998).

\bibitem{PhysRevLett.123.250401}
S.~Dutta and N.~R. Cooper, \enquote{{Critical Response of a Quantum van der Pol
  Oscillator},} {\protect\JournalTitle{Phys. Rev. Lett.}} \textbf{123}, 250401
  (2019).

\bibitem{gerry_knight_2004}
C.~Gerry and P.~Knight, \emph{{Introductory Quantum Optics}} (Cambridge
  University Press, 2004), chap. Field quantization, pp. 10--42.

\bibitem{10.1080}
A.~J. Daley, \enquote{Quantum trajectories and open many-body quantum systems,}
  {\protect\JournalTitle{Advances in Physics}} \textbf{63}, 77--149 (2014).

\bibitem{Rivas_2010}
\'{A}ngel Rivas, A.~D.~K. Plato, S.~F. Huelga, and M.~B. Plenio,
  \enquote{Markovian master equations: a critical study,}
  {\protect\JournalTitle{New Journal of Physics}} \textbf{12}, 113032 (2010).

\bibitem{Rivas_2014}
\'{A}ngel Rivas, S.~F. Huelga, and M.~B. Plenio, \enquote{{Quantum
  non-Markovianity: characterization, quantification and detection},}
  {\protect\JournalTitle{Reports on Progress in Physics}} \textbf{77}, 094001
  (2014).

\bibitem{PhysRevA.49.2771}
I.~Marzoli, J.~I. Cirac, R.~Blatt, and P.~Zoller, \enquote{{Laser cooling of
  trapped three-level ions: Designing two-level systems for sideband cooling},}
  {\protect\JournalTitle{Phys. Rev. A}} \textbf{49}, 2771--2779 (1994).

\bibitem{PhysRevA.46.2668}
J.~I. Cirac, R.~Blatt, P.~Zoller, and W.~D. Phillips, \enquote{Laser cooling of
  trapped ions in a standing wave,} {\protect\JournalTitle{Phys. Rev. A}}
  \textbf{46}, 2668--2681 (1992).

\bibitem{Carmichael1999}
H.~J. Carmichael, \emph{{Statistical Methods in Quantum Optics 1: Master
  Equations and Fokker-Planck Equations}} (Springer Berlin Heidelberg, 1999),
  chap. {Dissipation in Quantum Mechanics: The Master Equation Approach}, pp.
  "1--28".

\bibitem{RevModPhys.86.1391}
M.~Aspelmeyer, T.~J. Kippenberg, and F.~Marquardt, \enquote{Cavity
  optomechanics,} {\protect\JournalTitle{Rev. Mod. Phys.}} \textbf{86},
  1391--1452 (2014).

\bibitem{strogatz}
S.~H. Strogatz, \emph{{Nonlinear Dynamics and Chaos: With Applications to
  Physics, Biology, Chemistry, and Engineering}} (CRC press, 2015).

\bibitem{burden2011numerical}
R.~L. Burden and J.~D. Faires, \emph{Numerical Analysis} (Cengage Learning,
  Boston, MA, 2011).

\end{thebibliography}
	
\end{document}


\maketitle
	
	\section{Model Description}
	\label{section1}
	In a trapped-ion system, the ion has internal states and external vibrations, which can be coupled via interaction with the lasers. If the interaction only involves two internal levels and the vibration amplitude is small, it can be considered as a two-level system and the vibration can be described as a harmonic oscillator. So the free Hamiltonian of single ion is given by ($\hbar=1$) \cite{tuidao},
	\begin{equation}
		\hat{H}_0=\nu {{{\hat{a}}}^{\dagger }}\hat{a}+\frac{1}{2}{{\omega }_{0}}{{{\hat{\sigma }}}_{z}},
	\end{equation}
	where the first term describes the vibration of the ion and the second term denotes the ion's internal degrees of freedom. $\hat{a}  ({{\hat{a}}^{\dagger}})$ is the annihilation (creation) operator for the ion's vibration with frequency $\nu$, and $\hat{\sigma }_{z}$ is Pauli operator with $\omega_0=\omega_e-\omega_g$ denoting the transition frequency between the ion's ground state $ \left|  g \right\rangle$ and excited state $\left| e \right\rangle$. 
	
	In thermal equilibrium, the occupation probabilities of a two-level system follow the Boltzmann distribution \cite{laser}. Here, the frequency $\omega_{0}$ is within the optical frequency range, so the ion remains in the ground state even at room temperature, and the occupation probabilities can only be changed by optical pumping. The interaction between a traveling wave laser and the ion, within the context of semi-classical theory and under the dipole approximation, can be described by the following Hamiltonian \cite{tuidao}, 
	\begin{equation}
		\hat{H}_{\operatorname{int}}=\Omega\hat{\sigma}_{+}\left[e^{i\left(\vec{k} \cdot \vec{r}-\omega_L t+\phi\right)}+e^{-i\left(\vec{k} \cdot \vec{r}-\omega_L t+\phi\right)}\right]+h.c.,
		\label{eq2}
	\end{equation}
	where the coupling strength $\Omega$ is related to transition matrix $\vec{r}_{e g}=\left\langle e\left| {\vec{r}} \right|g \right\rangle$. $\omega_L$, $\vec{k}$, and $\phi$ are the frequency, wave-vector, and phase of the laser, respectively, and $\vec{r}$ is the position vector of the ion. 
	$\vec{k} \cdot \vec{r}$ can be expressed in terms of creation and annihilation operators as $\vec{k} \cdot \vec{r}=\eta\left(\hat{a}^{\dagger}+\hat{a}\right)$, where $\eta=k cos \phi \sqrt{\hbar/2m\nu}$ is defined as the Lamb-Dicke parameter. In the Lamb-Dicke regime ($\eta^2(2 n+1) \ll 1$, $n$ is vibrational quantum number) \cite{tuidao}, Eq. \ref{eq2} can be simplified as ($\phi=0$), 
	\begin{equation}
		{{{\hat{H}}}_{\operatorname{int}}}=\Omega{{{\hat{\sigma }}}_{+}}\left\{ {{e}^{-i {{\omega }_{L}}t}}\left[1+i \eta\left(\hat{a}^{\dagger}+\hat{a}\right)\right]+{{e}^{i{{\omega }_{L}}t }}\left[1-i \eta\left(\hat{a}^{\dagger}+\hat{a}\right)\right] \right\}+h.c., 
	\end{equation}
	
	Besides, to discuss the quantum synchronization of the phonon laser, the vibrational mode can be driven by an external field with driving strength $F$ and frequency $\omega_d$, which is represented as \cite{3981,PhysRevLett.123.250401},
	\begin{equation}
		{{{\hat{H}}}_{\text{ext}}}=iF\left( {{{\hat{a}}}^{\dagger }}{{e}^{-i\omega_{d} t}}-\hat{a}{{e}^{i\omega_{d} t}} \right).
	\end{equation}
	
	It is convenient for calculation by transforming into a rotating frame, where the total Hamiltonian $\hat{H}=\hat{H}_0+\hat{H}_{\text{int}}+\hat{H}_{\text{ext}}$ has no explicit time dependence. The unitary transformation is chosen as $U\left( t \right)=\exp \left\{-it\left[\left({{\omega }_{L}}-\omega_{d} \right){{\hat{\sigma }}_{z}}/2+\omega_{d} {{\hat{a}}^{\dagger }}\hat{a}\right]\right\}$. Similar to optical lasers, the coherent generation of phonons is essential for phonon lasers, which can be achieved through the first blue sideband resonance, where the laser frequency is set as $\omega_{L}=\omega_0 +\nu$. Using the Baker–Hausdorf lemma \cite{gerry_knight_2004},
	\begin{equation}
		{{e}^{\xi A}}B{{e}^{-\xi A}}=B+\xi \left[ A,B \right]+\frac{{{\xi }^{2}}}{2!}\left[ A,\left[ A,B \right] \right]+\ldots \ldots, 
	\end{equation}
	the operators after the unitary transformation take the following form, 
	\begin{equation}
		\begin{aligned}
			U^{\dagger}\hat{a}U&=\hat{a}e^{-i\omega_d t},\\
			U^{\dagger}\hat{a}^{\dagger}U&=\hat{a}^{\dagger}e^{i\omega_d t},\\
			U^{\dagger}\hat{\sigma}_{-}U&=\hat{\sigma}_{-}e^{-i(\omega_L-\omega_{d}) t}\\
			U^{\dagger}\hat{\sigma}_{+}U&=\hat{\sigma}_{+}e^{i(\omega_L-\omega_{d}) t},\\
		\end{aligned}
	\end{equation}
	and the time-independent Hamiltonian reads,
	\begin{equation}
		\begin{aligned}
			\hat{H}_{\text{I}}&=i\frac{{\partial U^{\dagger}} }{{\partial t} } U+{U}^{\dagger} \hat{H} {U}\\
			&=-\omega_d\hat{a}^\dagger\hat{a}-\frac{1}{2}\left(\omega_{L}-\omega_{d}\right)\hat{\sigma}_z+U^{\dagger}\left(\hat{H}_0+\hat{H}_{\text{int}}+\hat{H}_{\text{ext}}\right)U\\
			&=\left(\nu-\omega_{d}\right)\hat{a}^{\dagger}\hat{a}+\frac{1}{2}\left(\omega_{0}+\omega_{d}-\omega_{L}\right)\hat{\sigma }_{z}+U^{\dagger}\left(\hat{H}_{\text{int}}+\hat{H}_{\text{ext}}\right)U\\
			&=\left(\nu-\omega_{d}\right)\hat{a}^{\dagger}\hat{a}+\frac{1}{2}\left(\omega_{d}-\nu \right) \hat{\sigma }_{z}+iF\left( {{{\hat{a}}}^{\dagger }}-\hat{a} \right)+U^{\dagger}\hat{H}_{\text{int}}U,
		\end{aligned}
	\end{equation}
	with 
	\begin{equation}
		\begin{aligned}
			U^{\dagger}\hat{H}_{\text{int}}U=&\Omega e^{-i \omega_L t}\left\{U^{\dagger}\hat{\sigma}_{+}\left[1+i \eta\left(\hat{a}^{\dagger}+\hat{a}\right)\right]U\right\}+\Omega e^{i \omega_L t}\left\{U^{\dagger}\hat{\sigma}_{+}\left[1-i \eta\left(\hat{a}^{\dagger}+\hat{a}\right)\right]U\right\}+h.c.\\
			=&\Omega{{e}^{-i{{\omega }_{L}}t}}{{{\hat{\sigma }}}_{+}}{{e}^{i\left( {{\omega }_{L}}-{{\omega }_{d}} \right)t}}\left( 1+i\eta {{{\hat{a}}}^{\dagger }}{{e}^{i{{\omega }_{d}}t}}+i\eta \hat{a}{{e}^{-i{{\omega }_{d}}t}} \right)\\
			&+\Omega{{e}^{i{{\omega }_{L}}t}}{{{\hat{\sigma }}}_{+}}{{e}^{i\left( {{\omega }_{L}}-{{\omega }_{d}} \right)t}}\left( 1-i\eta {{{\hat{a}}}^{\dagger }}{{e}^{i{{\omega }_{d}}t}}-i\eta \hat{a}{{e}^{-i{{\omega }_{d}}t}} \right) +h.c.\\ 
			=&\Omega{{{\hat{\sigma }}}_{+}}{{e}^{-i{{\omega }_{d}}t}}\left( 1+i\eta {{{\hat{a}}}^{\dagger }}{{e}^{i{{\omega }_{d}}t}}+i\eta \hat{a}{{e}^{-i{{\omega }_{d}}t}} \right)\\
			&+\Omega{{{\hat{\sigma }}}_{+}}{{e}^{i(2{{\omega }_{L}}-{{\omega }_{d}})t}}\left( 1-i\eta {{{\hat{a}}}^{\dagger }}{{e}^{i{{\omega }_{d}}t}}-i\eta \hat{a}{{e}^{-i{{\omega }_{d}}t}} \right)+h.c.\\
			\approx&i\eta \Omega \left({{\hat{a}}^{\dagger }}{{\hat{\sigma }}_{+}}-\hat{a}\hat{\sigma}_{-}\right),
			\label{eq8}
		\end{aligned}
	\end{equation}
	where all the fast oscillating terms have been neglected. Therefore, the total Hamiltonian in the rotating frame can be expressed as, 
	\begin{equation}
		{{\hat{H}}_{\text{I}}}=-\Delta {{\hat{a}}^{\dagger }}\hat{a}+\frac{1}{2}\Delta {{\hat{\sigma }}_{z}}+i\eta \Omega \left({{\hat{a}}^{\dagger }}{{\hat{\sigma }}_{+}}-\hat{a}\hat{\sigma}_{-}\right )+iF\left( {{{\hat{a}}}^{\dagger }}-\hat{a} \right),
		\label{eqs9}
	\end{equation}
	where the detuning is defined as $\Delta=\omega_d-\nu$. 
	\begin{figure}[ht!]
		\centering
		\includegraphics[width=8cm]{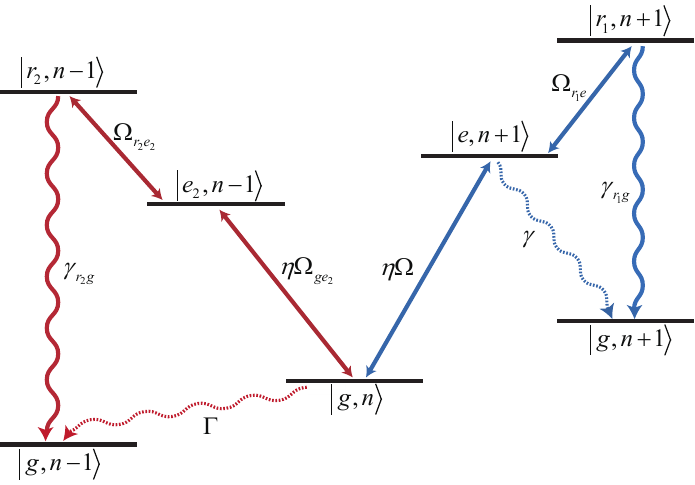}
		\caption{A sketch for the processes of generating phonon laser with a single trapped ion. The process of sideband heating (blue sideband) and sideband cooling (red sideband) of the ion's vibrational motion are represented by the blue and red lines, respectively. All the energy levels are composite product states including the ion's internal and external states, where $|g\rangle$, $|e\rangle$, $|e_2\rangle$, and $|r_i\rangle$ ($i=1,2$) correspond to the internal ground state, excited states and auxiliary states, respectively, and $|n\rangle$ denotes the external vibrational Fock states with $n$ phonons. The excited states $|e\rangle$, $|e_2\rangle$ are metastable states, while the auxiliary states $|r_1\rangle$, $|r_2\rangle$ are short-lived states. All the transitions with solid straight arrows are driven resonantly by appropriate laser frequencies with the corresponding Rabi frequencies marked next to the arrows. The solid wavy arrows indicate the rapid decay process of the auxiliary states into the ground state with the decay rates marked next to the arrows. For the sideband cooling, both the auxiliary state $|r_2\rangle$ and excited state $|e_2\rangle$ can be adiabatically eliminated, resulting in a net effect of decreasing a phonon denoted by a red dashed curly arrow with an effective mechanical damping rate $\Gamma$. For the sideband heating, the auxiliary state $|r_1\rangle$ can be adiabatically eliminated, which is denoted by a blue dashed curly arrow to obtain an effective spontaneous emission rate $\gamma$ of the excited state $|e\rangle$ to the ground state $|g\rangle$.} 
		\label{Figs1}
	\end{figure}
	
	In order to generate a phonon laser, additional dissipation mechanisms need to be considered, including effective spontaneous emission of the excited state at a rate $\gamma$ and mechanical damping at a rate $\Gamma$. Besides, for typical atomic or photonic setups, the environment relaxes to equilibrium on a timescale much faster than the system dynamics \cite{PhysRevLett.123.250401,10.1080}. Under this Markov approximation \cite{Rivas_2010,Rivas_2014}, the master equation is simplified to the Lindblad form,
	\begin{equation}
		\dot{\hat{\rho }}=-i\left[ {{{\hat{H}}}_\text{I}},\hat{\rho } \right]+\gamma \mathcal{D}\left[ {{{\hat{\sigma }}}_{-}} \right]\hat{\rho }+\Gamma\mathcal{D}[\hat{a}]\hat{\rho },
		\label{eqs10}
	\end{equation}
	which is a linear differential equations for the density matrix $\hat{\rho}$ with Lindblad operator $\mathcal{D}[\hat{O}]\hat{\rho }=\hat{O}\hat{\rho }{{\hat{O}}^{\dagger }}-\left\{ {{{\hat{O}}}^{\dagger }}\hat{O},\hat{\rho } \right\}/2$. 
	As shown in Fig. \ref{Figs1}, the process of the sideband heating of the ion's vibrational motion is represented by the blue lines, while that of the mechanical damping through sideband cooling is represented by the red lines. And the excited states for cooling and heating are metastable states in order to satisfy the sideband-resolved condition, i.e., the energy level linewidth is much smaller than the vibrational frequency $\nu$. To return to the ground state, auxiliary energy levels with short lifetimes are needed. The effective spontaneous emission of the excited state $\left|e\right\rangle$ presented in the second term of master equation can be realized through auxiliary energy level $\left|r_1\right\rangle$. Following excitation by the first blue sideband coupling, the ion must be reset to the ground state on timescales smaller than the oscillator decay rate, which can be achieved through the process that the laser couples the transition $\left|e\right\rangle\leftrightarrow\left|r_1\right\rangle$ and 'quenches' the state $\left|e\right\rangle$ by exciting to $\left|r_1\right\rangle$ from which it decays fast back to the ground state $\left|g\right\rangle$, providing an ‘effective’ radiative decay rate $\gamma=\Omega _{{{r}_{1}}{e}}^{2}/{{\gamma }_{{{r}_{1}}g}}$ for the excited state $\left|e\right\rangle$. And the condition ${{\Omega }_{{{r}_{1}}{{e}}}}\ll {{\gamma }_{{{r}_{1}}g}}$ must be satisfied to adiabatically eliminate the auxiliary level $\left|r_1\right\rangle$ \cite{PhysRevA.49.2771}. 
	The mechanical damping presented in the third term of master equation can be achieved by the sideband cooling \cite{PhysRevA.49.2771}. The cooling laser couples the transition $\left|g\right\rangle\leftrightarrow\left|e_2\right\rangle$ with the first red sideband resonance. After completing one round of cooling process, the ion is reset to the ground state $\left|g\right\rangle$ under the condition ${{\Omega }_{{{r}_{2}}{{e}_{2}}}}\ll {{\gamma }_{{{r}_{2}}g}}$, resulting in an effective spontaneous emission of $\left|e_2\right\rangle$ at a rate of $\Omega _{{{r}_{2}}{{e}_{2}}}^{2}/{{\gamma }_{{{r}_{2}}g}}$. The excited state $\left| {{e}_{2}}\right\rangle $ can be further adiabatically eliminated under the condition $\eta {{\Omega }_{g{{e}_{2}}}}\ll \Omega _{{{r}_{2}}{{e}_{2}}}^{2}/{{\gamma }_{{{r}_{2}}g}}$. Therefore, the cooling process is equivalent to mechanical damping at a rate of $\Gamma ={{\left( {\eta{\Omega }_{g{{e}_{2}}}}/{{\Omega }_{{{r}_{2}}e_2}} \right)}^{2}}{{\gamma }_{{{r}_{2}}g}}$ \cite{PhysRevA.46.2668,PhysRevA.49.2771}.
	
	\section{Derivation of the mean field approximation results}
	\label{section2}
	\subsection{phonon laser with external drive}
	\label{section21}
	Under the mean-field approximation, quantum operators can be separated into $\hat{O}=\left\langle\hat{O}\right\rangle+\delta\hat{O}$, where $\left\langle\hat{O}\right\rangle=\text{Tr}\left(\hat{O}\hat{\rho}\right)$ represents the mean-field results describing the classical behavior of the system, and $\delta\hat{O}$ represents the quantum fluctuation with zero mean value around the classical orbit. Here, we focus on the classical behaviors of the system, disregarding quantum noise terms and weak quantum correlations between the internal and external degrees of freedom of the ion, i.e., assuming that $\left\langle\hat{O}_1\hat{O}_2\right\rangle=\left\langle\hat{O}_1\right\rangle\left\langle\hat{O}_2\right\rangle$. Thus, the classical mean-field dynamical equations can be derived as follows,
	\begin{equation}
		\begin{aligned}
			\left\langle {\dot{\hat{a}}} \right\rangle =&\text{Tr}\left(\hat{a}\dot{\hat{\rho }}\right)\\
			=&i\Delta\text{Tr}\left( \hat{a}{\hat{a}}^{\dagger }\hat{a}\hat{\rho }-\hat{a}\hat{\rho }\hat{a}^{\dagger}\hat{a}\right)+\frac{1}{2}i\Delta\text{Tr}\left( \hat{a}\hat{\sigma}_z\hat{\rho }-\hat{a}\hat{\rho }\hat{\sigma}_z\right)+F\text{Tr}\left(\hat{a}\hat{a}^{\dagger }{\hat{\rho }}
			-\hat{a}\hat{a}\hat{\rho }-\hat{a}\hat{\rho}\hat{a}^{\dagger}+\hat{a}\hat{\rho}\hat{a}\right)\\
			&+\eta \Omega \text{Tr}\left[ \hat{a}({{{\hat{\sigma }}}_{+}}{{{\hat{a}}}^{\dagger }}\hat{\rho }-{{{\hat{\sigma }}}_{-}}\hat{a}\hat{\rho })-\hat{a}(\hat{\rho }{{{\hat{\sigma }}}_{+}}{{{\hat{a}}}^{\dagger }}-\hat{\rho }{{{\hat{\sigma }}}_{-}}\hat{a}) \right] \\ 
			& +\gamma \text{Tr}\left[ \hat{a}\left({{{\hat{\sigma }}}_{-}}\hat{\rho }{{{\hat{\sigma }}}_{+}}-\frac{1}{2}{{{\hat{\sigma }}}_{+}}{{{\hat{\sigma }}}_{-}}\hat{\rho }-\frac{1}{2}\hat{\rho }{{{\hat{\sigma }}}_{+}}{{{\hat{\sigma }}}_{-}}\right) \right] +\Gamma \text{Tr}\left[ \hat{a}(\hat{a}\hat{\rho }{{{\hat{a}}}^{\dagger }}-\frac{1}{2}{{{\hat{a}}}^{\dagger }}\hat{a}\hat{\rho }-\frac{1}{2}\hat{\rho }{{{\hat{a}}}^{\dagger }}\hat{a}) \right] \\ 
			=&i\Delta\text{Tr}\left[\left(\hat{a}\hat{a}^{\dagger}-\hat{a}^{\dagger}\hat{a}\right)\hat{a}\hat{\rho}\right]+F\text{Tr}\left[\left(\hat{a}\hat{a}^{\dagger}\hat{\rho}-\hat{a}\hat{\rho}\hat{a}^{\dagger}\right)\right]+\eta \Omega \text{Tr}\left[{{{\hat{\sigma }}}_{+}}\hat{\rho }\left(\hat{a}{{{\hat{a}}}^{\dagger }}-{{{\hat{a}}}^{\dagger }}\hat{a}\right)\right]\\
			&+\frac{\Gamma }{2}\text{Tr}\left[\left( {{{\hat{a}}}^{\dagger }}\hat{a}-\hat{a}{{{\hat{a}}}^{\dagger }} \right)\hat{a}\hat{\rho }\right]\\
			=& \eta \Omega \left\langle {{{\hat{\sigma }}}_{+}} \right\rangle - \left(\frac{\Gamma }{2}-i\Delta\right) \left\langle {\hat{a}} \right\rangle+F,  \\ 
			\left\langle\dot{\hat{\sigma}}_{+}\right\rangle  =&\eta \Omega \left\langle\hat{\sigma}_z\right\rangle\left\langle\hat{a}\right\rangle-\frac{\gamma}{2}\left\langle\hat{\sigma}_{+}\right\rangle, \\
			\left\langle\dot{\hat{\sigma}}_{z}\right\rangle  =&2 \eta \Omega\left(\left\langle\hat{\sigma}_{+}\right\rangle\left\langle\hat{a}^{\dagger}\right\rangle-\left\langle\hat{\sigma}_{-}\right\rangle\langle\hat{a}\rangle\right)-\gamma\left(\left\langle\hat{\sigma}_z\right\rangle+1\right).
			\label{eqs11}
		\end{aligned}
	\end{equation}
	The cyclic property of the trace $\text{Tr}({{\hat{O}}_{1}}{{\hat{O}}_{2}}{{\hat{O}}_{3}})=\text{Tr}({{\hat{O}}_{3}}{{\hat{O}}_{1}}{{\hat{O}}_{2}})=\text{Tr}({{\hat{O}}_{2}}{{\hat{O}}_{3}}{{\hat{O}}_{1}})$ \cite{Carmichael1999}, and the commutation relation $\left[ \hat{a},{{{\hat{a}}}^{\dagger }} \right]=1$, are utilized in the above calculation. Additionally, the variables $\langle\hat{a}\rangle=x_1+i y_1$, $\langle\hat{\sigma}_{-}\rangle=x_2+i y_2$, and $\langle\hat{\sigma}_z\rangle=z$ can be defined. Then, Eq. \ref{eqs11} can be written as,
	\begin{equation}
		\begin{aligned}
			{{{\dot{x}}}_{1}}&=\eta \Omega{{x}_{2}}-\frac{\Gamma}{2}{{x}_{1}}-\Delta{{y}_{1}}+F,\\
			{{{\dot{y}}}_{1}}&=\eta\Omega{{y}_{2}}-\frac{\Gamma}{2}{{y}_{1}}+\Delta{{x}_{1}},\\
			{{{\dot{x}}}_{2}}&=-\eta\Omega{{x}_{1}}z-\frac{\gamma}{2}{{x}_{2}}-\Delta{{y}_{2}},\\ 
			{{{\dot{y}}}_{2}}&=-\eta\Omega{{y}_{1}}z-\frac{\gamma}{2}{{y}_{2}}+\Delta{{x}_{2}},\\ 
			\dot{z}&=4\eta \Omega\left( {{x}_{1}}{{x}_{2}}+{{y}_{1}}{{y}_{2}}\right)-\gamma \left(z+1\right).
		\end{aligned}
		\label{eqs12}
	\end{equation}
	The nonlinearity in Eq. \ref{eqs12} originates from the fact that the Hamiltonian in Eq. \ref{eqs9} does not take the form of quadratic terms such as $\hat{a}\hat{a}, \hat{a}^{\dagger}\hat{a}^{\dagger}$, leading to various nonlinear dynamics \cite{RevModPhys.86.1391}. The first step is to analyze the fixed point solution $(x_{10},y_{10},x_{20},y_{20},z_{0})$, which can be obtained by setting the first derivatives in Eq. \ref{eqs12} to be zero. In parameter region B of Fig. 1(b) in the main text, there are three fixed points, while there is only one fixed point in other regions. The stability of these fixed points can be assessed by analyzing their corresponding Jacobian matrix, which is derived by linearizing Eq. \ref{eqs12} around each fixed point solution \cite{strogatz}. Hence, the Jacobian matrix is given by the following expression,
	\begin{equation}
		M=\left(\begin{array}{ccccc}
			-\frac{\Gamma}{2} & -\Delta& \eta \Omega & 0 & 0 \\
			\Delta & -\frac{\Gamma}{2} & 0 & \eta \Omega & 0 \\
			-\eta \Omega z_0 & 0 & -\frac{\gamma}{2} &-\Delta & -\eta \Omega x_{10} \\
			0 & -\eta \Omega z_{0} & \Delta & -\frac{\gamma}{2} & -\eta \Omega y_{10} \\
			4 \eta \Omega x_{20} & 4 \eta \Omega y_{20} & 4 \eta \Omega x_{10} & 4 \eta \Omega y_{10} & -\gamma
		\end{array}\right).
		\label{eqs13}
	\end{equation}
More specifically, a fixed point is considered stable if all eigenvalues of the Jacobian matrix have negative real parts. In the context of Fig. 1(b), the fixed point in region A is stable, while the fixed point in region C is unstable and the system will develop into limit cycle motion after some transient evolution. In region B, one of the three fixed points is stable, whereas the other two are unstable. In the frame rotating with the driving frequency, synchronization with the external drive implies the existence of a stable fixed point in the system. Therefore, region A and B demonstrate synchronization of phonon lasers with external drive in the absence of noise, while region C is characterized by unsynchronized behavior. Besides, the boundary between synchronization and unsynchronization regions shown in Fig. 1(b) can be obtained by the bisection method \cite{burden2011numerical}.
	\subsection{phonon laser without external drive}
\begin{figure}[ht!]
	\centering
	\includegraphics[width=10cm]{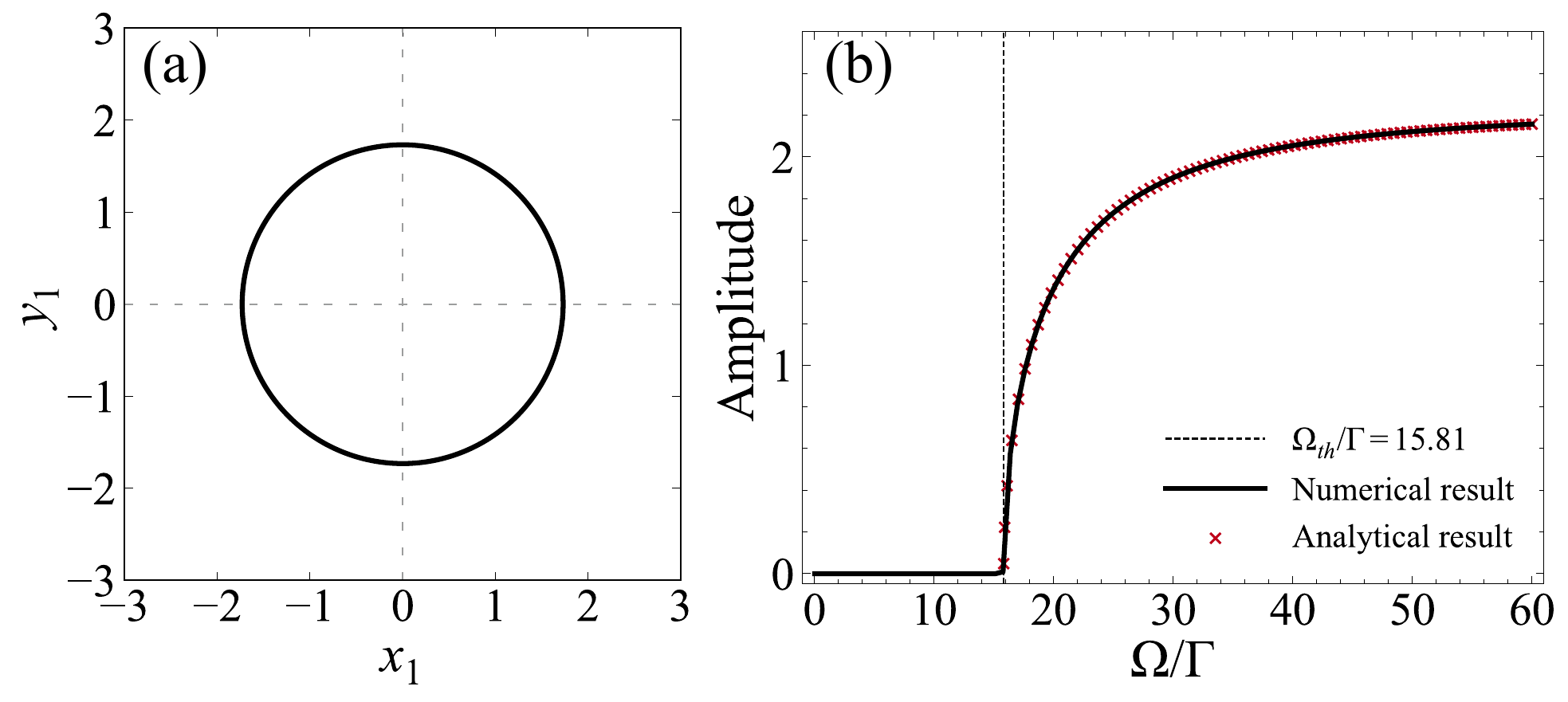}
	\caption{(a) The limit cycle under the mean-field approximation is centered at the origin with parameters $\eta=0.1, \Omega/\Gamma=25$, and $\gamma /\Gamma =10$. (b) Comparison of the amplitude of the limit cycle from numerical simulation of Eq. \ref{eqs15} and analytical expression $\sqrt{{\gamma}/(2\Gamma)-{\gamma^2}/{(8\eta^2\Omega^2)}}$. Below the threshold, there is only one stable fixed point $(x_{10}, y_{10}, x_{20}, y_{20}, z_0)=(0,0,0,0,-1)$, so the amplitude is zero. The parameters are $\eta=0.1, \gamma /\Gamma =10$. }
	\label{figS2}
\end{figure}

	In the absence of external drive, the discussion in part \ref{section21} is simplified for generating the single-ion phonon laser. The dynamics can be described by the following master equation by setting $F=0$, $\omega_{d}=0$ in Eqs. \ref{eqs9} and \ref{eqs10},
\begin{equation}
	\dot{\hat{\rho }}=-i\left[ \nu \hat{a}^{\dagger} \hat{a}-\frac{1}{2} \nu \hat{\sigma}_z+i \eta \Omega\left(\hat{\sigma}_{+} \hat{a}^{\dagger}-\hat{\sigma}_{-} \hat{a}\right),\hat{\rho }\right]+\gamma \mathcal{D}\left[ {{{\hat{\sigma }}}_{-}} \right]\hat{\rho }+\Gamma\mathcal{D}[\hat{a}]\hat{\rho }.
	\label{eqs14}
\end{equation}
Similarly, the classical mean-field dynamical equations are as follows, 
\begin{equation}
	\begin{aligned}
		\dot{x}_1 & =\eta \Omega x_2-\frac{\Gamma}{2} x_1+\nu y_1, \\
		\dot{y}_1 & =\eta \Omega y_2-\frac{\Gamma}{2} y_1-\nu  x_1, \\
		\dot{x}_2 & =-\eta \Omega x_1 z-\frac{\gamma}{2} x_2+\nu  y_2, \\
		\dot{y}_2 & =-\eta \Omega y_1 z-\frac{\gamma}{2} y_2-\nu  x_2, \\
		\dot{z} & =4 \eta \Omega\left(x_1 x_2+y_1 y_2\right)-\gamma(z+1).
	\end{aligned}
 \label{eqs15}
\end{equation}
The system has only one fixed point, given by $(x_{10}, y_{10}, x_{20}, y_{20}, z_0)=(0,0,0,0,-1)$. If this fixed point is unstable, the system will evolve into the limit cycle motion, indicating the generation of phonon laser. The eigenvalues of the corresponding Jacobian matrix can be obtained as follows,
	\begin{equation}
		\lambda=\left\{-\gamma, \frac{1}{4}\left(-\gamma-\Gamma \pm \sqrt{{\left(\gamma-\Gamma\right)}^2+16 \eta^2 \Omega^2}\pm4\nu i\right)\right\}.
	\end{equation}
	So the boundary of stability and instability for the fixed point is given by the equation, 
	\begin{equation}
		-\gamma-\Gamma \pm \sqrt{{\left(\gamma-\Gamma\right)}^2+16 \eta^2 \Omega^2}=0.
	\end{equation}
	In other words, the threshold condition for the phonon laser can be expressed as $\Omega_{\mathrm{th}}=\sqrt{\gamma \Gamma} / 2 \eta$, above which the fixed point is unstable, exhibiting a limit cycle motion oscillating with frequency $\nu$. As illustrated in Fig. \ref{figS2}(a), this limit cycle is centered at the origin and its amplitude can be solved by a smart method. Suppose you rotate with the same frequency $\nu$ with the limit cycle, then what you observe is a stable fixed point. Due to the arbitrariness of the initial phase, there exist numerous stable fixed points. Therefore, by transforming Eq. \ref{eqs14} into a new frame rotating with the vibrational frequency $\nu$, the original limit cycle attractor is translated into numerous stable fixed points distributed along the cycle. This can be accomplished by another unitary transformation $U^{\prime}\left( t \right)=\exp \left[-it\left(\nu\hat{a}^{\dagger}\hat{a}-\nu\hat{\sigma}_z/2\right)\right]$. Then the amplitude of the limit cycle can be determined by the coordinates of these newly developed stable fixed points. After the transformation, the master equation takes the following form,
\begin{equation}
	\dot{\hat{\rho }}=-i\left[i \eta \Omega\left(\hat{\sigma}_{+} \hat{a}^{\dagger}-\hat{\sigma}_{-} \hat{a}\right),\hat{\rho }\right]+\gamma \mathcal{D}\left[ {{{\hat{\sigma }}}_{-}} \right]\hat{\rho }+\Gamma\mathcal{D}\left[\hat{a}\right]\hat{\rho },
\end{equation}
 and the classical mean-field equations of the system reads,
	\begin{equation}
		\begin{aligned}
			{{{\dot{x}}}_{1}}&=\eta \Omega{{x}_{2}}-\frac{\Gamma}{2}{{x}_{1}},\\
			{{{\dot{y}}}_{1}}&=\eta\Omega{{y}_{2}}-\frac{\Gamma}{2}{{y}_{1}},\\
			{{{\dot{x}}}_{2}}&=-\eta\Omega{{x}_{1}}z-\frac{\gamma}{2}{{x}_{2}},\\ 
			{{{\dot{y}}}_{2}}&=-\eta\Omega{{y}_{1}}z-\frac{\gamma}{2}{{y}_{2}},\\ 
			\dot{z}&=4\eta \Omega\left( {{x}_{1}}{{x}_{2}}+{{y}_{1}}{{y}_{2}}\right)-\gamma \left(z+1\right).
		\end{aligned}
		\label{eqs19}
	\end{equation}
In the lasing regime, in addition to the original unstable fixed point $(x_{10}, y_{10}, x_{20}, y_{20}, z_0)=(0,0,0,0,-1)$, the newly developed stable fixed points satisfy the following expressions,
	\begin{equation}
		\begin{aligned}
			&x_{10}^2+y_{10}^2 =\frac{\gamma}{2 \Gamma}-\frac{\gamma^2}{8 \eta^2 \Omega^2}, \\
			&x_{20}  =\frac{\Gamma x_{10}}{2 \eta \Omega}, y_{20}=\frac{\Gamma y_{10}}{2 \eta \Omega}, z_0 =-\frac{\gamma \Gamma}{4 \eta^2 \Omega^2}.
		\end{aligned}
		\label{eqs20}
	\end{equation}
 As shown in Fig. \ref{figS2}(b), the amplitude of the limit cycle from numerical simulation of Eq. \ref{eqs15} is equal to the radius of the circle on which the newly developed stable fixed points $(x_{10},y_{10})$ in Eq. \ref{eqs20} are located, given by analytical expression $\sqrt{{\gamma}/(2\Gamma)-{\gamma^2}/{(8\eta^2\Omega^2)}}$. 

	\bibliography{samplesup}